\documentclass[english,12pt]{article}

\usepackage{setspace}
\usepackage[margin=1in]{geometry}

\usepackage{color,soul} 
\usepackage{amsmath}
\usepackage{graphicx} 
\usepackage{tabularx,multirow}
\usepackage{natbib}
\usepackage{amsfonts} 
\newcolumntype{Y}{>{\centering\arraybackslash}X}

\usepackage{titlesec}

\titleformat*{\section}{\large\bfseries}
\titleformat*{\subsection}{\bfseries}
\titleformat*{\subsubsection}{\bfseries}

\def\bA{\textbf{A}}

\def\bK{\textbf{K}}

\def\bE{\textbf{E}}
\def\bC{\textbf{C}}

\def\bY{\textbf{Y}}
\def\bZ{\textbf{Z}}
\def\ba{\textbf{a}}

\def\bc{\textbf{c}}
\def\be{\textbf{e}}

\def\bh{\textbf{h}}

\def\br{\textbf{r}}

\def\bz{\textbf{z}}
\def\cbar{\bar{\mathbf{c}}}
\def\bcbar{\bar{\mathbf{c}}}
\def\bCbar{\bar{\mathbf{C}}}
\def\E{\text{E}}

\def\exp{\text{exp}}

\def\bbeta{\boldsymbol{\beta}}
\def\balpha{\boldsymbol{\alpha}}
\def\btheta{\boldsymbol{\theta}}
\def\bgamma{\boldsymbol{\gamma}}

\def\bSigma{\boldsymbol{\Sigma}}

\def\bSig\mathbf{\Sigma}

\begin{document}

\thispagestyle{empty}

\begin{center}
{\bf \Large Bayesian kernel machine regression--causal mediation analysis}
\end{center}

\begin{center}
Katrina L. Devick\textsuperscript{1}, Jennifer F. Bobb\textsuperscript{2}, Maitreyi Mazumdar\textsuperscript{3,4}, Birgit Claus Henn\textsuperscript{5}, \\David C. Bellinger\textsuperscript{3,4}, David C. Christiani\textsuperscript{4}, Robert O. Wright\textsuperscript{6}, Paige L. Williams\textsuperscript{7,8}, \\ Brent A. Coull\textsuperscript{4,7}, and Linda Valeri\textsuperscript{9,*}
\end{center}

\vspace{1mm}

{\small
\noindent $^1$Department of Quantitative Health Sciences, Mayo Clinic, Scottsdale, AZ \\
$^2$Kaiser Permanente Washington Health Research Institute, Seattle, WA\\
$^3$Department of Neurology, Boston Children's Hospital, Boston, MA\\
$^4$Department of Environmental Health, Harvard T.H. Chan School of Public Health, Boston, MA \\
$^5$Department of Environmental Health, Boston University School of Public Health, Boston, MA\\
$^6$Department of Environmental Medicine and Public Health, Icahn School of Medicine at Mount Sinai, New York, NY \\
$^7$Department of Biostatistics, Harvard T. H. Chan School of Public Health, Boston, MA \\
$^8$Department of Epidemiology, Harvard T. H. Chan School of Public Health, Boston, MA \\
$^9$Department of Biostatistics, Columbia Mailman School of Public Health, New York, NY\\
$^*$email: lv2424@cumc.columbia.edu
}
\vspace*{5mm}


\noindent{\bf Summary:}
Greater understanding of the pathways through which an environmental mixture operates is important to design effective interventions. We present new methodology to estimate natural direct and indirect effects and controlled direct effects of a complex mixture exposure on an outcome through a mediator variable. We implement Bayesian Kernel Machine Regression (BKMR) to allow for all possible interactions and nonlinear effects of (1) the co-exposures on the mediator, (2) the co-exposures and mediator on the outcome, and (3) selected covariates on the mediator and/or outcome. From the posterior predictive distributions of the mediator and outcome, we simulate counterfactuals to obtain posterior samples, estimates, and credible intervals of the mediation effects. Our simulation study demonstrates that when the exposure-mediator and exposure-mediator-outcome relationships are complex, BKMR--Causal Mediation Analysis performs better than current mediation methods. We applied our methodology to quantify the contribution of birth length as a mediator between \emph{in utero} co-exposure to arsenic, manganese and lead, and children's neurodevelopmental scores, in a prospective birth cohort in Bangladesh. Among younger children, we found a negative (adverse) association between the metal mixture and neurodevelopment. We also found evidence that birth length mediates the effect of exposure to the metal mixture on neurodevelopment for younger children. If birth length were fixed to its $75^{th}$ percentile value, the harmful effect of the metal mixture on neurodevelopment is attenuated, suggesting nutritional interventions to help increase fetal growth, and thus birth length, could potentially block the harmful effect of the metal mixture on neurodevelopment.

\vspace*{7mm}\noindent{\bf Key words}: Children's neurodevelopment; Environmental mixture; Mixture; Multi-pollutant exposure.
\thispagestyle{empty}

\newpage
\clearpage
\setcounter{page}{1}

\section{Introduction}
\label{P2:intro}

The ability to identify pathways through which a complex exposure mixture operates is critical for the development of public health policy, as mediation pathways can often be influenced via interventions. In addition, exposure to environmental mixtures, as opposed to independently acting single agents, represents the real-life exposure scenario. Therefore, the National Institute of Environmental Health Sciences (NIEHS) has prioritized the development of statistical methods that quantify the effect of environmental mixtures on health outcomes.\citep{carlin2013, braun2016, taylor2016} With this increased priority, new methodology is needed to measure mixtures effects and the pathways by which they operate if we are to minimize the burden of disease. 

Exposure to complex mixtures is ubiquitous, and recent work in toxicology and epidemiology now emphasizes assessing mixtures of chemicals. Since elements of a mixture have the potential to exhibit complex interactions, it is important to consider the whole mixture when evaluating the nature of the relationship of multiple chemicals on a health outcome.\citep{wright2006, claushenn2012, claushenn2014} Once a relationship between a mixture and outcome is established, questions regarding the pathways, or mechanisms, through which the mixture operates arise. Of note, in our examples, mechanisms can be biological processes (inflammation, endocrine disruption, etc.) or can be markers of biological processes (height, weight, etc.). While we recognize that these markers are not explicit mechanisms, they often are useful as they represent composites of several biological mechanisms. For example, height (or birth length) reflects fetal nutrition, placental health and function, inflammation, and likely genetics.  We use the term “mechanism” when referring to birth length with the understanding that it is a marker of multiple biological mechanisms and not a mechanism itself. It should have all the properties of a biological mediator and may yield higher power to detect mediated effects since it reflects the sum of multiple mechanisms. 

One approach to quantifying biological mechanisms is the use of causal mediation analysis.\citep{pearl2001,vanderweele2009, vanderweele2010,valeri2013} Causal mediation analysis allows for the decomposition of a total effect (TE) of an exposure on an outcome into the pathway that operates indirectly through an intermediate (mediator) variable and the pathway that is independent of the intermediate variable, or that operates directly from the exposure to the outcome. Researchers' understanding of the pathways operating through intermediate variables may inform policy recommendations to reduce the harmful impact of environmental mixtures on health outcomes. For example, interventions that increase birth length (which can be monitored in populations) may  include nutrition, inflammation reduction or prenatal care that improves placental function. The evidence that birth length mediates the effect of prenatal metal exposure on neurodevelopment neurodevelopment would point to this factor as a way to measure the benefit of interventions on neurodevelopment within populations at birth. Monitoring infant anthropometry, such as birth weight and birth length, may serve as an early sentinel for possible neurodevelopment effects at a later age.

Few methods exist to estimate mediation effects when the exposure of interest is a mixture and exposure-mediator interactions are present. Current methods require specifying an outcome and a mediator regression. If the mediator variable is restricted to have a linear effect on the outcome, closed form solutions are available to estimate the natural direct effect (NDE), natural indirect effect (NIE), and controlled direct effects (CDEs) of a mediator on the relationship of a mixture on an outcome.\citep{vanderweele2009,valeri2013} In the presence of a nonlinear effect of the mediator on the outcome, the algorithm presented by Imai, Keele, and Tingley can be used to estimate the NDE, NIE, and CDEs through prediction of counterfactuals.\citep{Imai2010} However, both of these available methods require the researcher's \emph{a priori} specification of all data complexities and assume no model misspecification. Thus, all true existing interactions between the individual elements of the mixture, the elements of the exposure mixture and mediator, and any nonlinearities need to be included in the models for the mediator and outcome to obtain unbiased estimates. As the dimensions of a multi-dimensional exposure increase, it becomes exponentially difficult to use current methods to obtain unbiased estimates of the mediated effects. To our knowledge, no other methods using a causal inference framework currently exist to estimate the NDE, NIE, and CDEs of a potentially complex exposure mixture on an outcome through a mediator variable.

In this paper, we present a novel causal mediation method to estimate the NDE, NIE, and CDEs for a potentially complex mixture of exposures on an outcome operating through an intermediate variable. We allow for highly complex exposure-mediator and exposure-mediator-response functions, in addition to complex exposure-covariate and exposure-mediator-covariate relationships with the mediator and outcome respectively, using Bayesian Kernel Machine Regression (BKMR). BKMR has been shown to perform well relative to other nonparametric techniques and simpler statistical approaches for estimating mixture effects.\citep{bobb2015} BKMR allows for variable selection, which has not been incorporated in other nonparametric models and is an important feature when the number of mixture elements is large and/or they are highly correlated, as in children's health research. Additionally, we use BKMR for the mediator and outcome regression models since BKMR allows for all possible nonlinearities and interactions among the elements included in the kernel with the specified outcome, without \emph{a priori} specification, and credible intervals obtained from a BKMR fit inherently control for multiple testing due to the Bayesian nature of the model and the prior specification.\citep{scott2010,bobb2018}

We predict counterfactuals using the posterior predictive distributions of the mediator and the outcome and present an algorithm for estimation of mediation effects. We also conduct a simulation study to compare how our approach performs relative to current mediation methods that assume a restrictive linear relationship between the exposure, mediator, and outcome. 

We apply our method to analyze data from a prospective birth cohort in Bangladesh. Arsenic, manganese, and lead are known neurotoxicants\citep{bressler1999,clarkson1987,polanska2013,vahter2008,zoni2013,lucchini2017} that are abundant in the Bangladeshi environment, including in drinking water,\citep{Kile2009} and the relationship between arsenic, manganese and lead on child neurodevelopment has been shown to be complex.\citep{wasserman2004,wasserman2006,wasserman2007,wasserman2008,wright2006,claushenn2010,claushenn2012,claushenn2014,hamadani2011,valeri2017} In our data application, we estimate the NDE and NIE to bring light to the relationship of this metal mixture on child neurodevelopment operating through \emph{in utero} growth, specifically birth length. Also, we estimate the CDE of the metal mixture on neurodevelopment at different quantiles of birth length to assess if adequate \emph{in utero} growth could potentially block some of the harmful effects of the metal mixture on neurodevelopment. Since the persistence of the effect of \emph{in utero} metal exposure is unknown, that is, younger children might exhibit stronger associations with \emph{in utero} metal exposure than older children, we estimate these effects for different aged children.


\section{Materials and Methods}
\label{P2:methods}

\subsection{Bayesian kernel machine regression}
We first review BKMR presented by Bobb et al. as a framework to estimate the effect of a complex mixture on a health outcome.\citep{bobb2015} For each subject $i=1,\dots,n$, we assume: 
\begin{equation}\label{bkmrmodel}
Y_i = h\left(\bZ_i\right) + \bC_i^T\bbeta + \epsilon_i,
\end{equation}
where $Y_i$ is a continuous health outcome, $\bZ_i=\left(Z_{1i},\dots,Z_{Li}\right)^T$ is a vector of length $L$ containing continuous exposure variables (e.g. metals) and other continuous variables that may have a complex relationship with these exposures and outcome (e.g. effect modifiers), $\bC_i=\left(C_{1i},\dots,C_{Pi}\right)^T$ is a vector of additional covariates assumed to have a linear effect on the outcome, and $\epsilon_i \overset{iid}{\sim} N(0,\sigma^2)$. Model (\ref{bkmrmodel}) relates the outcome to the exposure mixture through a flexible function, $h(\cdot)$, which accommodates nonlinear associations and/or interactions among the variables in $\bZ$ with the outcome $Y$. 

Identifying a set of basis functions to represent $h(\cdot)$ can be difficult, thus, we employ a kernel machine representation.\citep{cristianini2000} The unknown function $h(\cdot)$ can be specified in two ways. One can either use basis functions or a positive-definite kernel function $K(\cdot,\cdot)$ to identify $h(\cdot)$. Mercer's theorem \citep{cristianini2000} established that the kernel function $K(\cdot,\cdot)$ implicitly specifies a unique function space spanned by a particular set of orthogonal basis functions, under regularity conditions. Therefore, any $h(\cdot)$ in this function space can be represented by a set of basis functions or by the dual representation kernel function $K(\cdot,\cdot)$.  Liu, Lin and Ghosh showed that model (\ref{bkmrmodel}) can be expressed as the mixed model (\ref{bkmrmix})\citep{liu2007}:
\begin{align}\label{bkmrmix}
\begin{split}
Y_i ~&\sim~ N(h_i+ \bC_i^T\bbeta, \sigma^2)~~~\text{independent, } i=1,\dots n,\\
\bh ~&=~ (h_i, \dots, h_n)^T ~\sim~ N(0, \tau\bK),
\end{split}
\end{align}
where $\bK$, the kernel matrix, has $(i,j)$-th element $K(\bZ_i,\bZ_j)$.

The kernel function $K(\cdot,\cdot)$ uses a metric of similarity to establish how close exposure profiles $\bZ_i$ and $\bZ_j$ are for subjects $i$ and $j$. We will focus on the Gaussian kernel, which uses Euclidean distance as a means to quantify this similarity. Under the Gaussian kernel, we assume $\text{corr}(h_i,h_j)=\exp\left\{-(1/\rho)\sum_{\ell=1}^L(z_{i\ell}-z_{j\ell})^2\right\}$, where $\rho$ is a tuning parameter that regulates the smoothness of the dose-response function. Intuitively, this assumption means subjects with similar exposure profiles ($\bZ_i$ close to $\bZ_j$) will have more similar risks ($h_i$ will be close to $h_j$).

When the exposure mixture contains more than a few variables, (\ref{bkmrmodel}) can be fit with component-wise variable selection to add an additional layer of shrinkage above and beyond the smoothing induced by (\ref{bkmrmix}). This results in a more parsimonious representation of the mixture effect. To allow for variable selection within a Bayesian paradigm, the kernel function $K(\cdot, \cdot)$ is augmented. In the case of the Gaussian kernel, the kernel function is expanded as:
\begin{equation}
 K(\bZ_i, \bZ_j; \br) = \exp\left\{-\sum_{\ell=1}^Lr_{\ell}(Z_{i\ell}-Z_{j\ell})^2\right\},
 \end{equation}
 where $\br=(r_1, \dots, r_{\ell})^T$, and we assume a "spike-and-slab" prior for the auxiliary parameters, 
\begin{align}
\begin{split}
r_{\ell} | \delta_{\ell} ~&\sim~ \delta_{\ell}f_1(r_{\ell}) + (1-\delta_{\ell})P_0, ~~~\ell=1, \dots, L, \\
\delta_{\ell}~&\sim~\text{Bernoulli}(\pi),
\end{split}
\end{align}
where $\delta_{\ell}$ is an indicator that element $\ell$ is included in the kernel, $f_1(\cdot)$ denotes a probability density function with support on $\mathbb{R}^+$, and $P_0$ is the density with a point mass at 0. 
To fit (\ref{bkmrmodel}), we use a Markov chain Monte Carlo (MCMC) sampler. 
For additional details regarding BKMR, see Bobb et al.\citep{bobb2015, bobb2018}


\subsection{Causal mediation analysis}
In order to define causal contrasts in a mediation context, we first define our notation. Let $Y_{am}$ denote the counterfactual outcome $Y$ if the exposure $A$ was set to level $a$ and mediator $M$ was set to level $m$. Let $M_a$ be the counterfactual mediator $M$ that would have been observed if the exposure $A$ was set to level $a$. Accordingly, $Y_{aM_{a^*}}$ represents the counterfactual outcome $Y$ if the exposure $A$ was set to level $a$ and the mediator $M$ was set to the level it would have taken if the exposure $A$ was set to level $a^*$.
			
The mediated effects of interest, the total effect (TE), the natural direct effect (NDE), the natural indirect effect (NIE), and the controlled direct effects (CDEs), are formally defined as:
\begin{eqnarray}	
\text{TE}&=& \E\left[Y_{a}-Y_{a^*} \right], \\
\text{NDE}&=& \E\left[Y_{aM_{a^*}}-Y_{a^*M_{a^*}} \right], \\			
\text{NIE} &=& \E\left[Y_{aM_{a}}-Y_{aM_{a^*}}\right], \\
\text{CDE}(m)&=& \E\left[Y_{am}-Y_{a^*m} \right].		
\end{eqnarray}
The NDE captures the average difference in the counterfactual outcomes for a change in exposure level $a^*$ to $a$, while fixing the mediator to the level it would have taken if the exposure was set to $a^*$. The NIE measures the average difference in counterfactual outcomes when fixing the exposure to level $a$, while the mediator varies from the level it would have taken if the exposure was set to $a$ compared to $a^*$. The CDE quantifies the average difference in the counterfactual outcomes for a change in exposure level from $a^*$ to $a$, while intervening to fix the mediator to a specified level, $m$.


\subsection{Bayesian kernel machine regression--causal mediation analysis}
For BKMR--Causal Mediation Analysis (BKMR-CMA), we consider a single normally distributed health outcome $Y$, single normally distributed mediator variable $M$, continuous exposure mixture $\bA$, continuous effect modifiers $\bE_M$ that may have a complex relationship with the exposures on the mediator, and continuous effect modifiers $\bE_Y$ that may have a complex relationship with the exposures and mediator on the outcome. Let $\bZ_M=\left(\bA, \bE_M \right)$ and $\bZ_Y=\left(\bA, \bE_Y \right)$ be matrices containing all of the exposure variables (e.g. metals) and the effect modifiers to include in the prospective kernel functions. Although $\bZ_M$ and $\bZ_Y$ might be the same sets of variables, our framework allows for them to differ. Additionally, one or both of the sets of variables $\bE_M$ or $\bE_Y$ could be empty (i.e. there are no effect modifiers that interact with the exposures and/or have a nonlinear relationship with the mediator or outcome). Lastly, let $\bC$ be a matrix of confounders assumed to have a linear effect on the outcome. To allow for potentially complex relationships between the elements in $\bZ_M$, we model the mediator variable using the BMKR model (\ref{P2:medmodel}):
\begin{equation}\label{P2:medmodel}
	M_i = h_M(\bZ_{Mi}) + \bC_i^T\bbeta + \epsilon_{Mi}, 
\end{equation}
where $\epsilon_{Mi} \overset{iid}{\sim}N(0,\sigma_{M}^2)$. Since accounting for exposure-mediator interactions is important to obtain unbiased effect estimates, we include the mediator variable along with $\bZ_Y$ in the kernel function when modeling the health outcome in (\ref{P2:outmodel}):
\begin{equation}\label{P2:outmodel}
	Y_i = h_Y(\bZ_{Yi},M_i) + \bC_i^T\btheta + \epsilon_{Yi}, 
\end{equation}
where $\epsilon_{Yi} \overset{iid}{\sim}N(0,\sigma_{Y}^2)$. 
By fitting the models separately, we assume $\epsilon_{Mi}$ and $\epsilon_{Yi}$ are independent. To model the total effect of the exposure mixture on the outcome, we consider BKMR model (\ref{P2:TEmodel}):
\begin{equation}\label{P2:TEmodel}
	Y_i = h_{TE}(\bZ_{Yi}) + \bC_i^T\bgamma+\epsilon_{TEi}, 
\end{equation}
where $\epsilon_{TEi} \overset{iid}{\sim}N(0,\sigma_{TE}^2)$. 

We estimate the TE, NDE, and NIE using BKMR-CMA at particular levels of the effect modifiers, $\be_M$ and $\be_Y$, for a change in exposure mixture from $\ba^*$ to $\ba$ via the following algorithm, where we let $\bz_M=\left(\ba,\be_M\right)$, $\bz_M^*=\left(\ba^*,\be_M\right)$, $\bz_Y=\left(\ba,\be_Y\right)$, and $\bz_Y^*=\left(\ba^*,\be_Y\right)$. This corresponds to a fixed intervention in the exposures. It is of utmost importance that careful thought is given to the choice of values for $\ba^*$ and $\ba$ as well as $\be_M$ and $\be_Y$ used to estimate the mediation effects. These values should lie within the support of the multivariate distribution of the data to ensure the values compared are observable in the population under study\citep{zigler2021}. 

\begin{enumerate}
	\item Fit BKMR mediator, outcome, and total effect models (\ref{P2:medmodel}), (\ref{P2:outmodel}), and (\ref{P2:TEmodel}), respectively.		
			
	\item For each MCMC iteration, $j=1,\hdots, J $:
			
	\begin{enumerate}
		\item Generate $k=1, \hdots, K$ samples of the mediator for the mean level of covariates under exposure level $\ba^*$ from mediator model (\ref{P2:medmodel}), fixing the modifiers $\bE_M$ to level $e_M$: 
			\begin{eqnarray*}
			M_{\bz^*}^{(jk)}(\cbar) &=& \E^{(j)}(M | \bZ_M = \bz_M^*, \bC = \cbar) + \sigma_{M}^{(j)}N(0,1)\\
			 &=&  h_M^{(j)}(\bZ_M=\bz_M^*) + \cbar^T\bbeta^{(j)} + \sigma_{M}^{(j)}N(0,1).
			 \end{eqnarray*}
				
		\item  For each of the $j=1,\hdots,J$ iterations and $k=1,\hdots,K$ samples of $M_{\bz^*}$, jointly estimate the average outcome value for the mean level of covariates for $Y_{\bz M_{\bz^*}}$ and $Y_{\bz^* M_{\bz^*}}$ from outcome model (\ref{P2:outmodel}), setting the modifiers $\bE_Y$ to level $\be_Y$ and where $\tilde{\bz}=\begin{pmatrix} \bz_Y & \bz_Y^* \end{pmatrix}^T$:
		\begin{eqnarray*}
					Y_{\tilde{\bz} M_{\bz^*}}^{(jk)}(\cbar) &=& \E^{(j)}(Y | \bZ_Y = \tilde{\bz}, M= M_{\bz^*}^{(jk)}(\cbar), \bC=\cbar)\\
			& =& h_Y^{(j)}(\bZ_Y=\tilde{\bz},M=M_{\bz^*}^{(jk)}(\cbar)) + \cbar^T\btheta^{(j)}.
			\end{eqnarray*}
		
		\item Let the $j^{th}$ posterior sample of $Y_{\bz M_{\bz^*}}$ be the mean over the $K$ samples of $Y_{\bz M_{\bz^*}}$, or $Y_{\bz M_{\bz^*}}^{(j)}(\cbar)=\frac{1}{K} \sum_{k=1}^K Y_{\bz M_{\bz^*}}^{(jk)}(\cbar)$, and let the $j^{th}$ posterior sample of $Y_{\bz^* M_{\bz^*}}$ be the mean over the $K$ samples of $Y_{\bz^* M_{\bz^*}}$, or $Y_{\bz^* M_{\bz^*}}^{(j)}(\cbar)=\frac{1}{K} \sum_{k=1}^K Y_{\bz^* M_{\bz^*}}^{(jk)}(\cbar)$.

		\item We sample the $j^{th}$ posterior sample of $Y_{\bz^*}$ and $Y_{\bz}$ from the total effect model (\ref{P2:TEmodel}). We then calculate the average outcome value for the mean level of covariates at $\tilde{\bz}=\begin{pmatrix} \bz_Y & \bz_Y^* \end{pmatrix}^T$ from (\ref{P2:TEmodel}), setting the modifiers to level $\be_Y$: 
		 \begin{eqnarray*}
			Y_{\tilde{\bz}}^{(j)}(\cbar) &=& \E^{(j)}(Y | \bZ_Y = \tilde{\bz}, \bC = \cbar) \\
			 &=&  h_{TE}^{(j)}(\bZ_Y =\tilde{\bz}) + \cbar^T\bgamma^{(j)}
			 \end{eqnarray*}			
			\end{enumerate}

	\item Obtain the $j^{th}$ posterior sample of the TE, NDE, and NIE by:							\begin{eqnarray*}
				\text{TE}^{(j)} &=&  Y_{\bz}^{(j)}(\cbar)- Y_{\bz^*}^{(j)}(\cbar),\\
				\text{NDE}^{(j)} &=&  Y_{\bz M_{\bz^*}}^{(j)}(\cbar)- Y_{\bz^* M_{\bz^*}}^{(j)}(\cbar),\\
				\text{NIE}^{(j)}  &=&   \text{TE}^{(j)} - \text{NDE}^{(j)}.
			\end{eqnarray*}	
		Note that we estimate the NIE as the difference between the TE and NDE to ensure the property of effect decomposition and to minimize the potential effect of overfitting using kernel methods.	
	\item Estimate the TE, NDE, and NIE and 95\% credible intervals (CI) conditional for level of the effect modifiers and marginally according to the confounders as the posterior mean and posterior $2.5^{th}$ and $97.5^{th}$ percentiles. We take the standard Bayesian approach to calculate the CIs and variance of each effect of interest. After we calculate the marginal TE, NDE, and NIE at each MCMC iteration, we then calculate the variance and 2.5$^{th}$ and 97.5$^{th}$ percentiles of these $J$ estimates of the TE, NDE, and NIE separately, and use these as the variance estimate and 95\% CI. 
	\end{enumerate}

\noindent Four no unmeasured confounding assumptions are required for the NDE and NIE to have a causal interpretation: (i) $Y_{\ba m} \amalg \bA | \bC$, (ii) $Y_{\ba m} \amalg M | \bC, \bA$, (iii) $M_{\ba} \amalg \bA | \bC$, and (iv) $Y_{\ba m} \amalg M_{\ba^*} | \bC$. Namely, there are no unmeasured exposure-outcome confounders, there are no unmeasured mediator-outcome confounders, there are no unmeasured exposure-mediator confounders, and the exposure mixture does not affect any mediator-outcome confounders.\citep{vanderweele2010} To estimate CDEs, only two no unmeasured confounding assumptions are required: (i) and (ii). The algorithm to estimate CDEs is similar to the algorithm presented above. We include explicit steps to estimate CDEs in Appendix \ref{P2:CDEalgorithm}. 

We define BKMR-CMA-VS to be the method in which BKMR models (\ref{P2:medmodel}), (\ref{P2:outmodel}), and (\ref{P2:TEmodel}) are all fit with component-wise variable selection and then these models are used in our BKMR-CMA algorithm to obtain posterior samples of the counterfactuals and mediation effects. R Code to implement BKMR-CMA and BKMR-CMA-VS is publicly available on GitHub (https://github.com/kdevick/BKMR-CMA). 



\section{Simulation Study}
\label{P2:sim}

We evaluated the ability of BKMR-CMA and BKMR-CMA-VS to estimate the joint mediated effects of a mixture compared to the product method \citep{baron1986} and causal mediation analysis methods using linear models \citep{vanderweele2009, vanderweele2010, valeri2013} under numerous plausible data generating mechanisms.

\subsection{Setup}
We first generated a true underlying dataset for each simulation scenario. The dataset consisted of a continuous health outcome $Y_i$, a continuous mediator value $M_i$, and an exposure mixture $\bA_i=(A_{1i}, \dots,A_{Li})^T$ comprised of $L$ continuous mixture components for $i=1,\dots,~1,000,000$ subjects. The exposure mixture was generated as $\bA_i \sim \mathcal{N}(\bf{0},\bSigma)$, the mediator as $M_i\sim N(h_M(A_{i1}, \dots,A_{iQ}),$ $\sigma_{M}^2)$, and the health outcome as  $y_i\sim N(h_Y(A_{i1}, \dots,A_{iR},M_i),$ $\sigma_{Y}^2)$, where we assumed that the mediator depended on subset $Q<L$ of the exposure mixture components and the health outcome depended on the exposure subset $R<L$ in addition to the mediator. We set $\sigma_M^2=\sigma_Y^2=0.75$, motivated from the signal-to-noise ratios observed in the Bangladesh application (Section \ref{s:P2data}). 

\begin{figure}[h]
  \centering
  \includegraphics[width=2in]{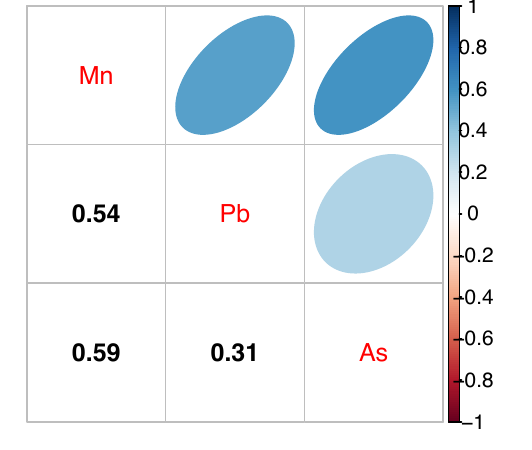}
\includegraphics[width=3in]{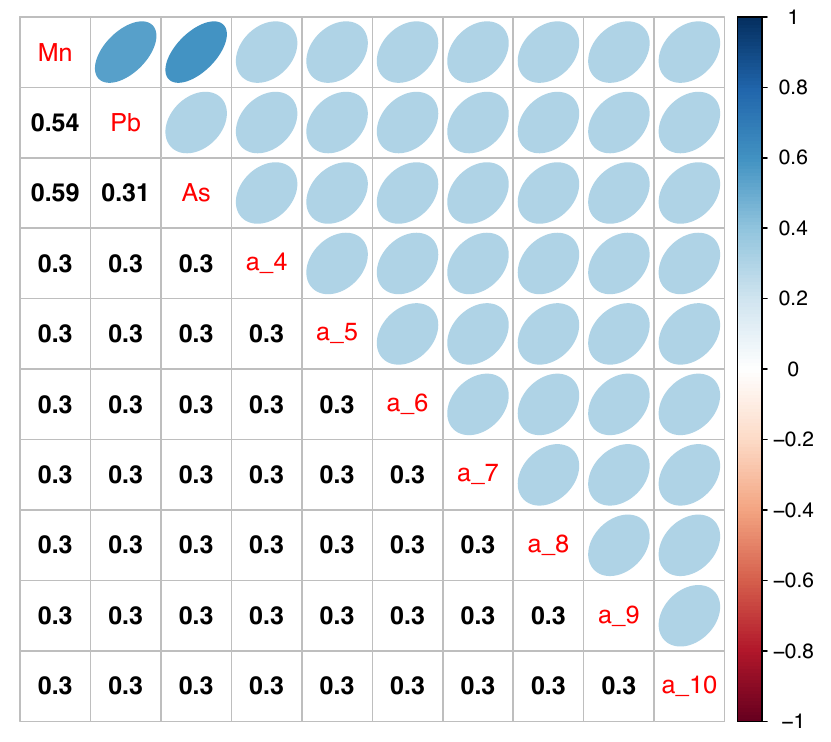}\\
 \caption[Covariance structure considered in our simulation when $L=3$ and $L=10$. The covariance for manganese (Mn), arsenic (As), and lead (Pb) from Bangladesh after log transformation and standardization]{Covariance structure $\boldsymbol{\Sigma}$ considered in our simulation when $L=3$ and $L=10$. The covariance for manganese (Mn), arsenic (As), and lead (Pb) from Bangladesh after log transformation and standardization.} 
 \label{corr}
\end{figure}

We considered two cases for the number of mixture components, $L=3$ and $L=10$. For $L=3$, we took the true covariance matrix $\bSigma$ to be the covariance structure of \emph{in utero} exposure to manganese (Mn), arsenic (As), and lead (Pb) after log transformation and standardization from our Bangladesh application.  For $L=10$, we used the covariance structure from the Bangladesh data for the first three elements of the mixture, and then set the remaining covariances to 0.3. 
The correlation structures used in our simulation are summarized in Figure \ref{corr}.

\begin{figure}[h]
  \centering
\includegraphics[width=2in]{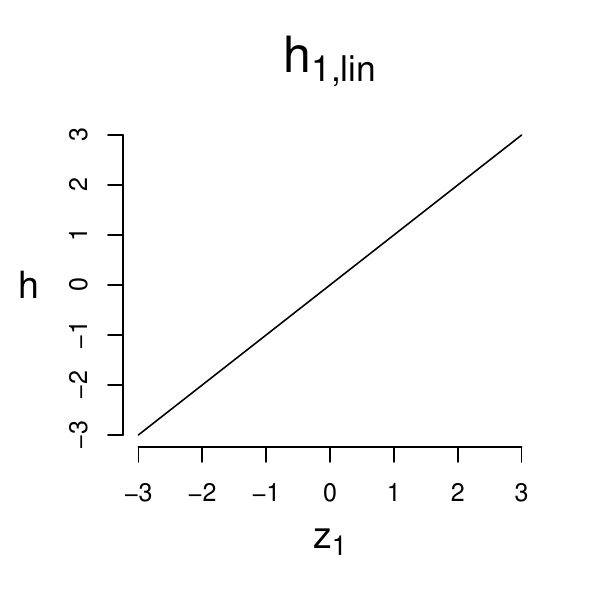}
\includegraphics[width=2in]{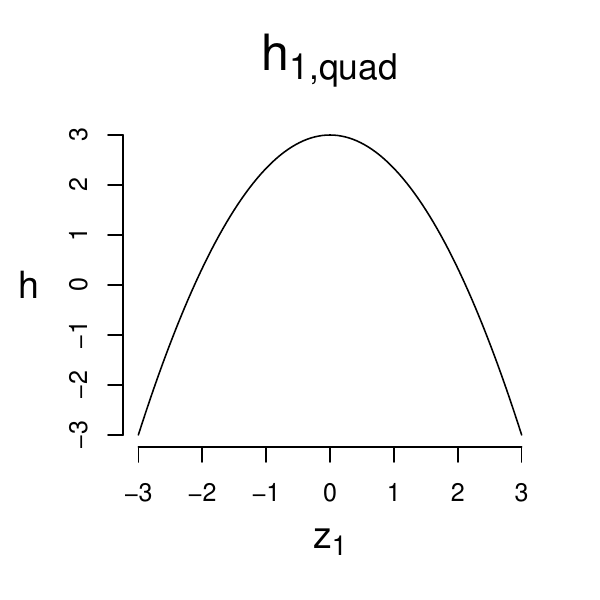}\\

\vspace*{-1mm}\hspace*{1mm}\includegraphics[width=2.25in]{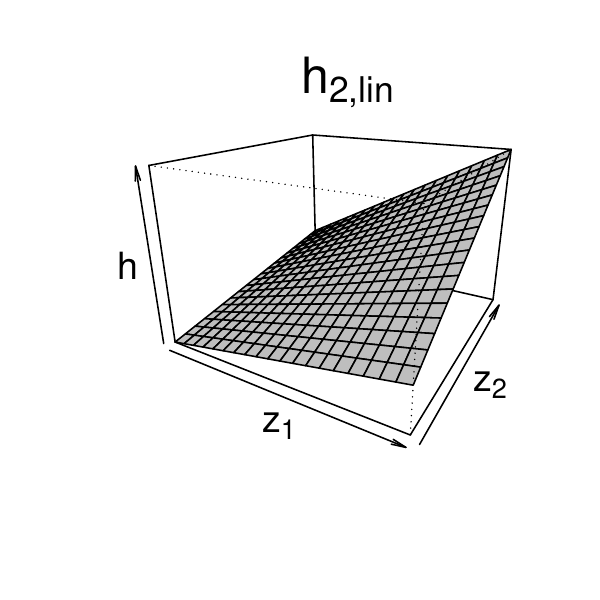}
\hspace*{-4mm}\includegraphics[width=2.25in]{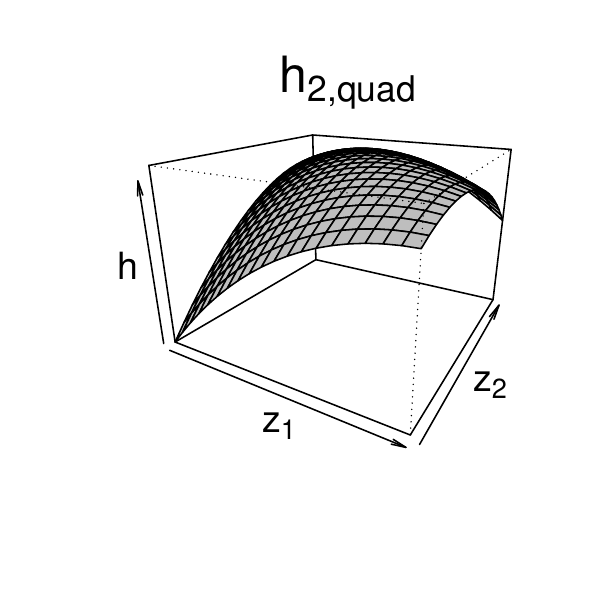}\\
 \caption[Dose-response surfaces used in our simulation to generate the mediator and outcome values.]{ Dose-response surfaces used in our simulation to generate the mediator and outcome values. $h_{2,lin}$ and $h_{2,quad}$ include an additive interaction of the 2 components, $z_1$ and $z_2$. The specific functions we used are the following: 
 $h_{1,lin}(a_1) = a_1$, 
$h_{1,quad}(a_1) = - \frac{2a_1^2}{3} + 3$,
$h_{2,lin}(a_1,a_2)=\frac{a_1}{2}+\frac{a_2}{2}+\frac{a_1a_2}{12} - \frac{3}{4}$, and 
$h_{2,quad}(a_1,a_2)= -\frac{(a_1-1)^2}{4}-\frac{(a_2-1)^2}{4}-\frac{(a_1+3)(a_2+3)}{3}+5$. }
 \label{hfunc}
\end{figure}

We considered six scenarios for the true underlying exposure-mediator and exposure-response functions $h_M(\cdot)$ and $h_Y(\cdot)$, respectively. Figure \ref{hfunc} graphically displays the functions we considered for $h_M(\cdot)$ and $h_Y(\cdot)$ in our simulations. The specifications for each scenario, including the metals considered to have an effect on the mediator and outcome and the functions $h_M(\cdot)$ and $h_Y(\cdot)$ used to generate the true underlying dataset, are summarized in Table \ref{simscenarios}. The linear method is correctly specified for scenario 1 as $h_M(\cdot)$ and $h_Y(\cdot)$ are linear functions. Since all scenarios include an interaction term, the traditional method is misspecified for all scenarios. 
Our first scenario included a null effect of all metals on the mediator and an additive interaction between one metal, $A_1$, and the mediator on the outcome. In this scenario 1,  a linear model with an additive interaction is the "truth". We also considered numerous nonlinear scenarios where the functions $h_M(\cdot)$ and $h_Y(\cdot)$ are quadratic. For scenarios 1 through 3, the true NIE is null, and for scenario 4, the true NDE is null. For each simulation scenario, we randomly sampled 500 datasets of 300 observations each, $\{Y_i, \bA_i, M_i\}_{i=1}^{300}$, from the true underlying dataset of one million subjects.

For each simulation dataset, we first fit BKMR models (\ref{modelsim}) - (\ref{modelsim2}) with and without component-wise variable selection: 
\begin{eqnarray}\label{modelsim}
	M_i &=& \beta_0 + h_M(\bA_i) + \epsilon_{Mi} ,\\ 
	Y_i &=& \theta_0 + h_Y(\bA_i,M_i) +\epsilon_{Yi} , \\ \label{modelsim2}
	Y_i &=& \gamma_0 + h_{TE}(\bA_i) +\epsilon_{TEi} , 
\end{eqnarray}
where $\epsilon_{Mi}  \overset{iid}{\sim}N(0,\sigma_{{M} }^2)$, $\epsilon_{Yi}  \overset{iid}{\sim}N(0,\sigma_{{Y} }^2)$, and $\epsilon_{TEi} \overset{iid}{\sim}N(0,\sigma_{TE}^2)$. Using our proposed BKMR-CMA and BKMR-CMA-VS approaches, we estimated the NDE, NIE, and TE for a change of the exposure mixture from $\ba^*$, the exposures set equal to their $25^{th}$ percentile in the true underlying dataset, to $\ba$, where the exposures were set equal to their $50^{th}$ percentile in the true underlying dataset. Since the exposures were generated from a multivariate normal distribution with mean 0 and variance 1, all elements in the $\ba^*$ vectors were extremely close to $\Phi^{-1}\left(0.25\right)=-0.674$ (range: [-0.677,-0.672]) and all elements in the $\ba$ vectors were extremely close to $\Phi^{-1}\left(0.5\right)=0$ (range: [-0.001,0.0008]). We estimated the CDE for the same change in the exposures, from $\ba^*$ to $\ba$, fixing the mediator at the observed $25^{th}$, $50^{th}$, and $75^{th}$ percentiles of the mediator in the corresponding true underling dataset. 

Second, we conducted mediation analyses for the joint effect of the metal mixture using both causal mediation methods and traditional approaches. Specifically, we used causal mediation methods or the "linear approach" in which we fit linear regression models for both the mediator and outcome and allowed for exposure-mediator interactions in the outcome model.\citep{vanderweele2009,vanderweele2010,valeri2013} For the "traditional approach," we estimated the NDE and NIE without considering exposure-mediator interactions using the product method extended for multiple exposures.\citep{baron1986} Derivations of the closed form solutions to estimate causal mediation effects under the linear and traditional methods are includedin Appendix \ref{s:causalmedformulas}. For these analyses, we considered the same change in exposures as with BKMR-CMA and BKMR-CMA-VS, a change in the metals from their $25^{th}$ to $50^{th}$ percentiles in the underlying data for the NDE, NIE, TE, and CDEs. For the CDEs, we fixed the mediator to its $25^{th}$, $50^{th}$, and $75^{th}$ percentiles in the corresponding true underling dataset.  

\subsection{Simulation results}

\begin{figure}[h]
\hspace*{7.5mm}\includegraphics[height=2.2in]{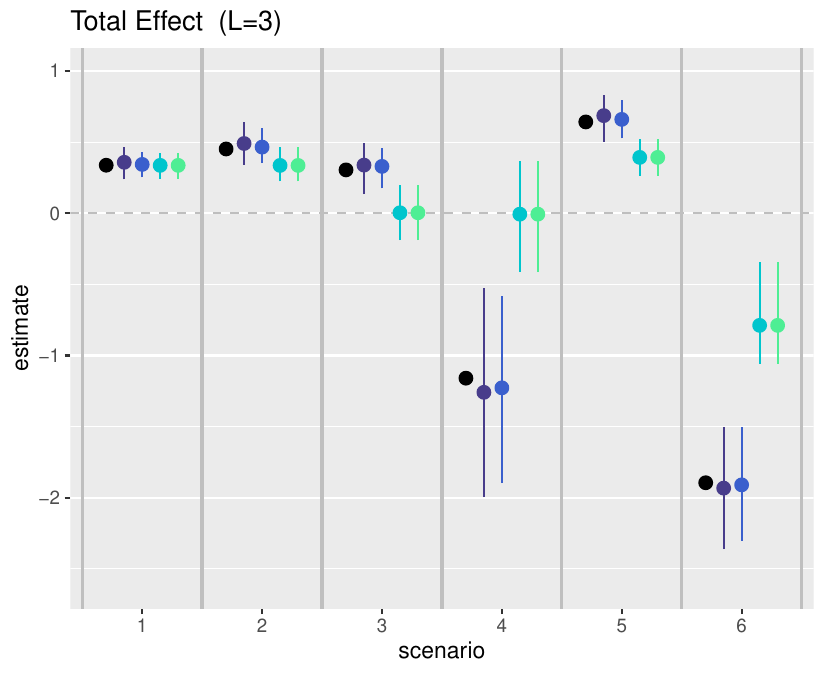}
\includegraphics[height=2.2in]{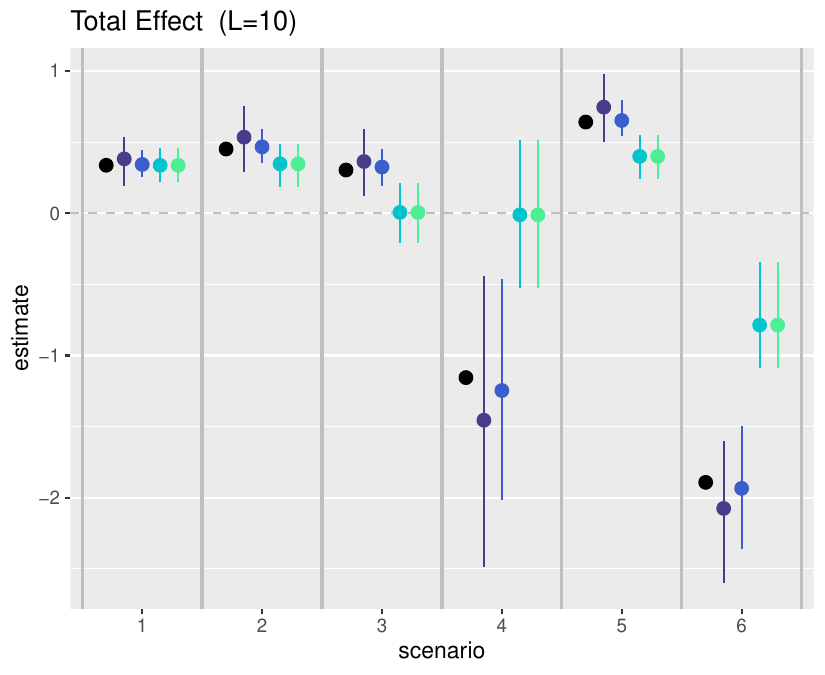}\\

\vspace*{-2mm}\hspace*{7.5mm}\includegraphics[height=2.2in]{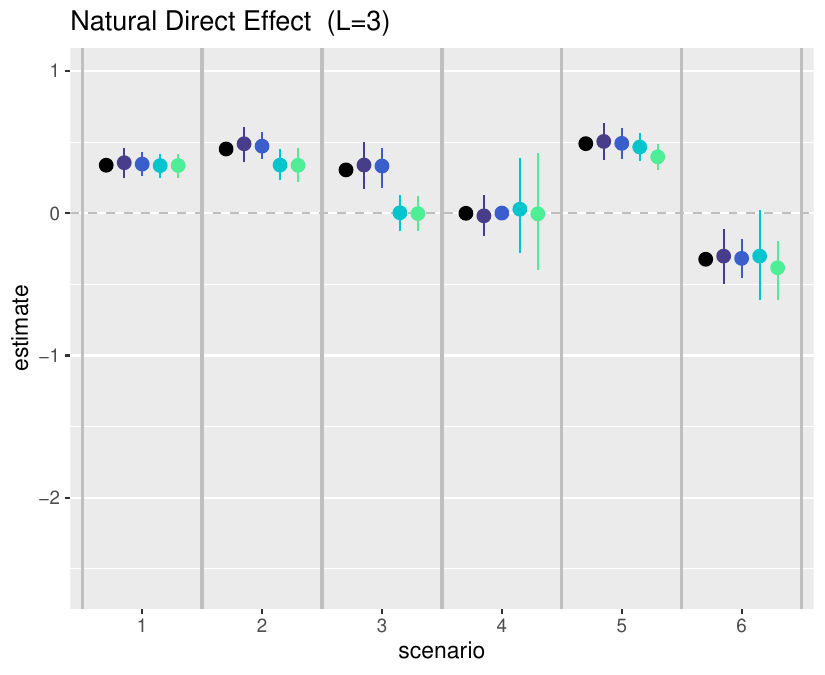}
\includegraphics[height=2.2in]{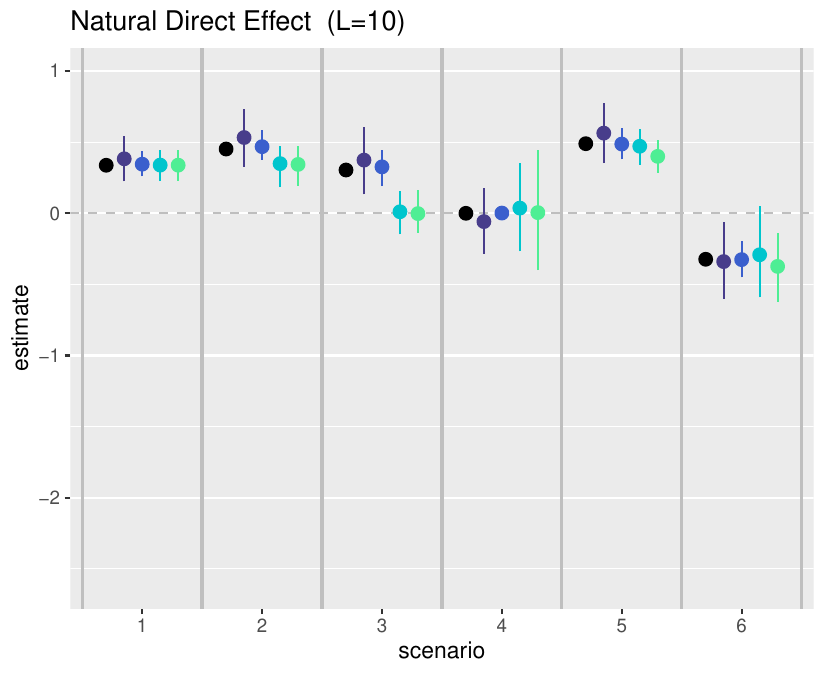}\\

\vspace*{-2mm}\hspace*{7.5mm}\includegraphics[height=2.2in]{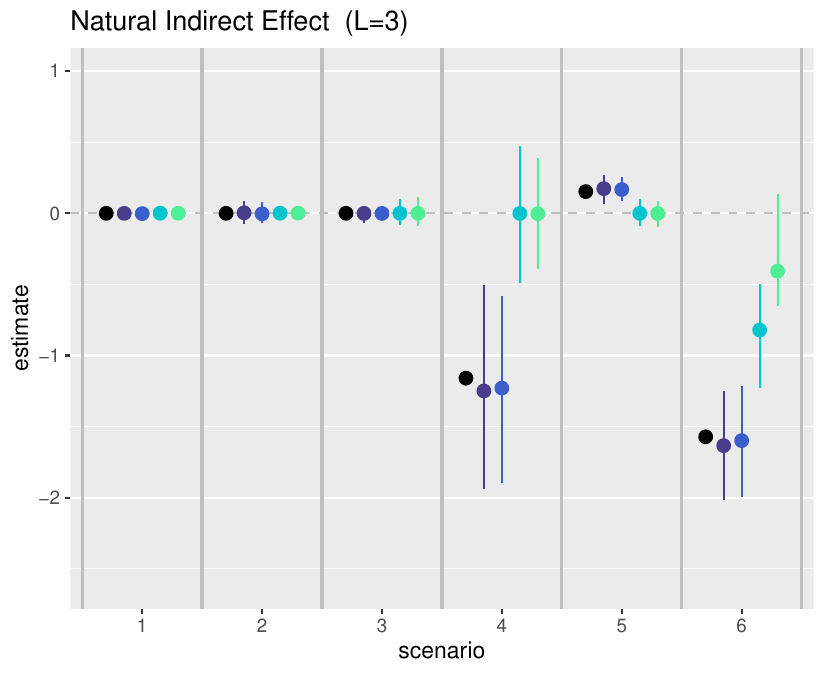}
\includegraphics[height=2.2in]{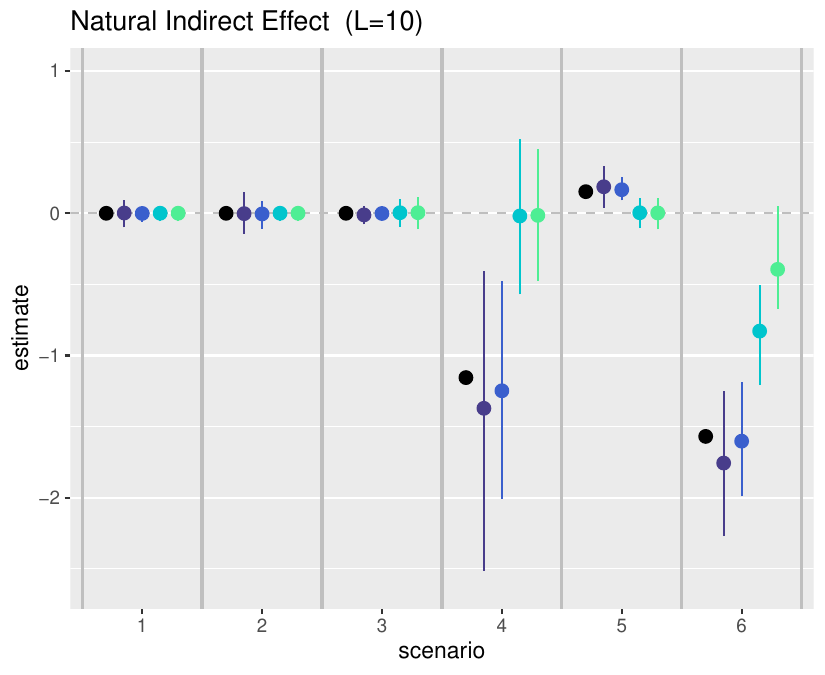}
\includegraphics[height=2.2in]{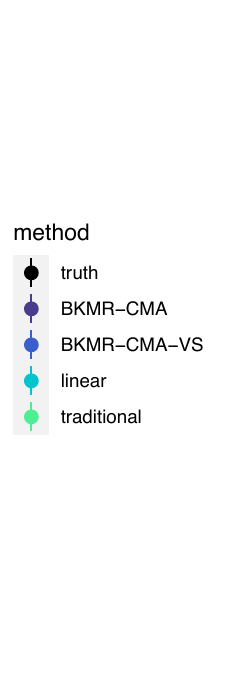}

 \caption[Simulation results]{Empirical median and $2.5^{th}$ and $97.5^{th}$ percentiles calculated from the estimates of the TE, NDE and NIE across the 500 simulation datasets using our proposed BKMR-CMA and BKMR-CMA-VS approaches, the linear method, and the traditional method under each simulation scenario. The specific data generation functions used for each simulation scenario are defined in Table \ref{simscenarios}. The truth for each mediation effect and scenario are depicted as black dots. Results are shown for six different data generation scenarios and when the number of mixture components is three and ten. } 
 \label{P2simresults}
\end{figure}

\begin{figure}[h]
\hspace*{7.5mm}\includegraphics[height=2.2in]{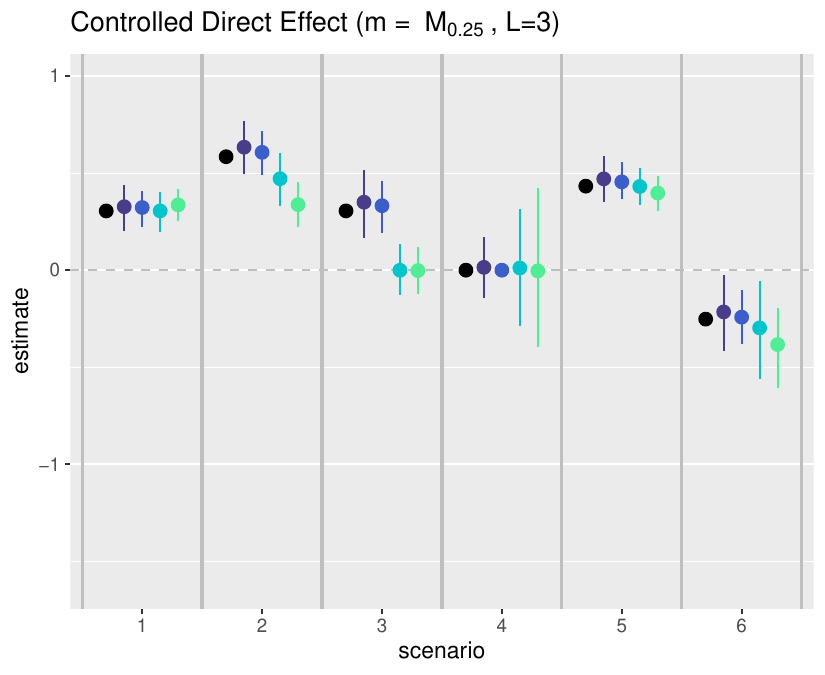}
\includegraphics[height=2.2in]{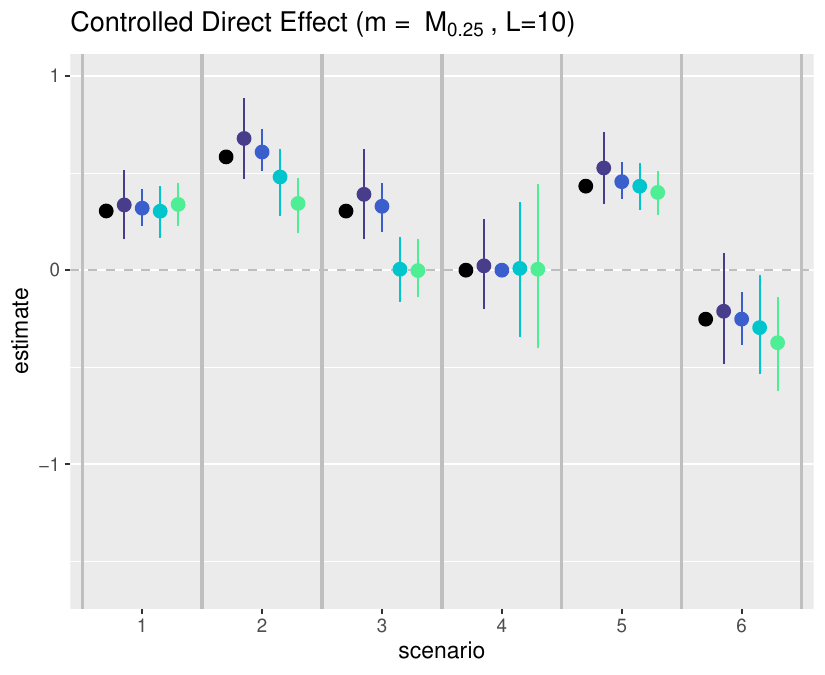}\\

\vspace*{-2mm}\hspace*{7.5mm}\includegraphics[height=2.2in]{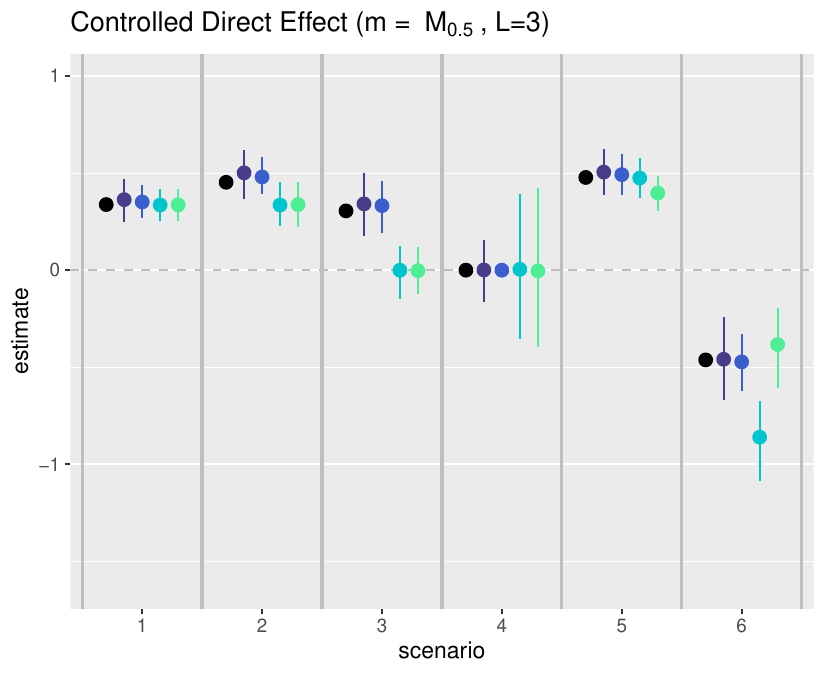}
\includegraphics[height=2.2in]{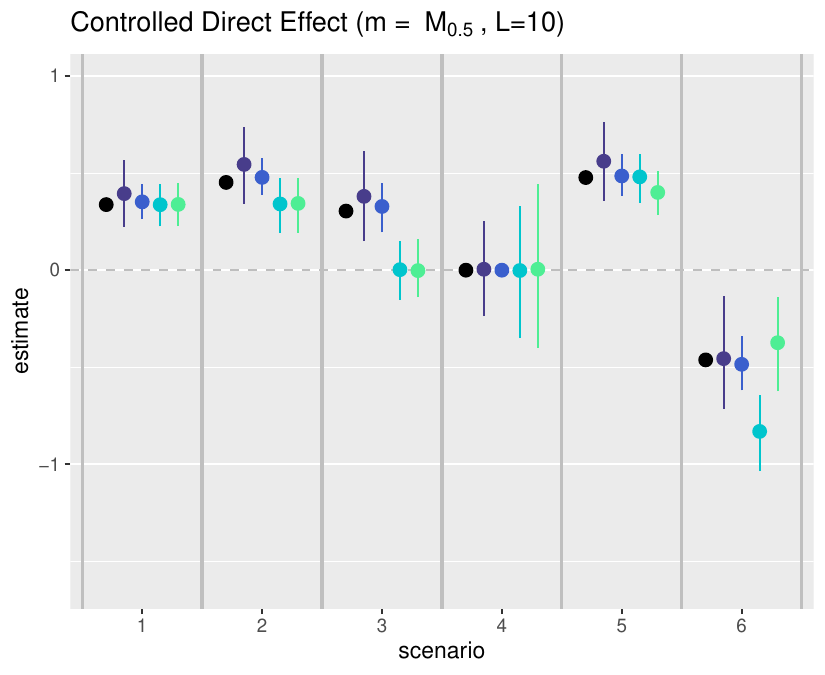}\\

\vspace*{-2mm}\hspace*{7.5mm}\includegraphics[height=2.2in]{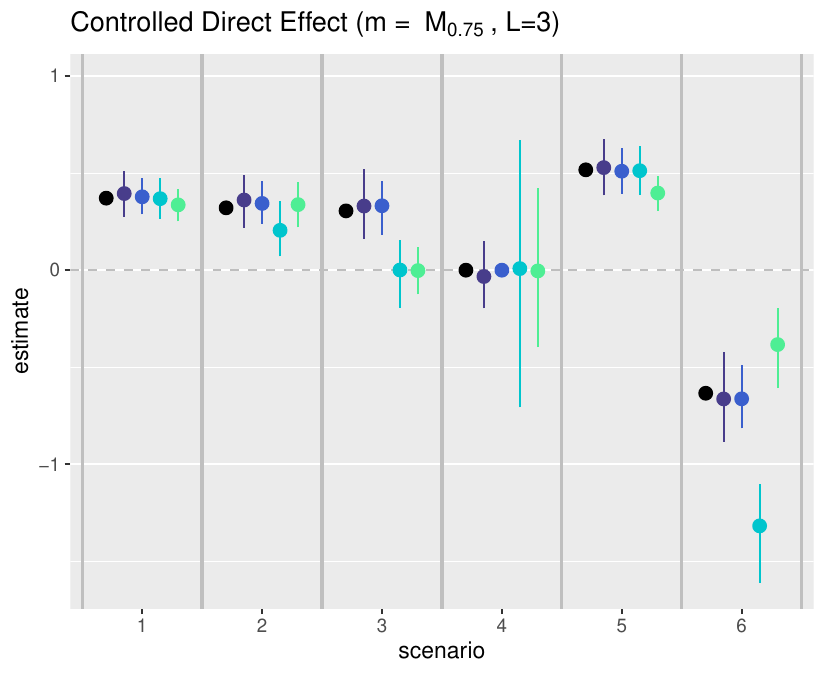}
\includegraphics[height=2.2in]{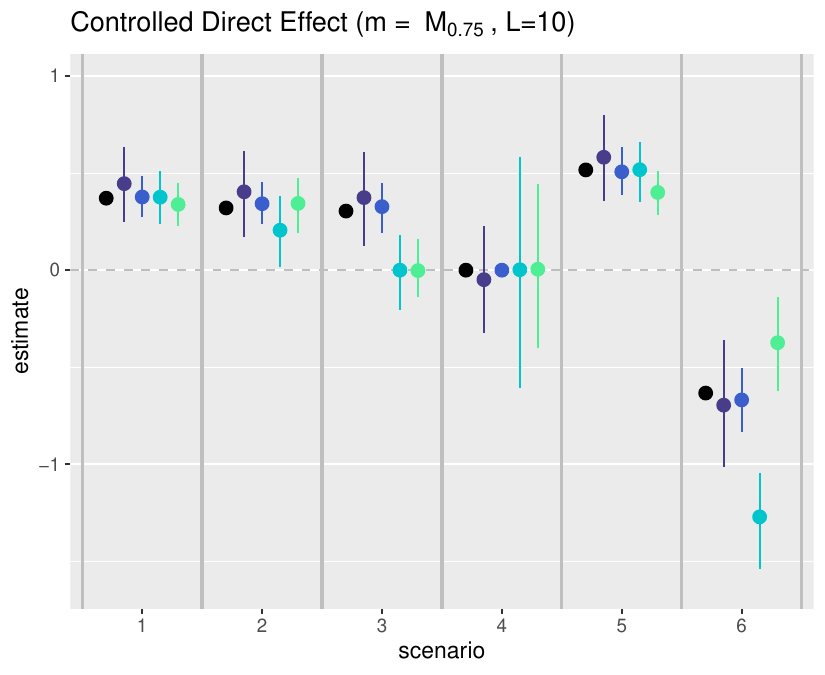}
\includegraphics[height=2.2in]{P2simlegend}

 \caption[Simulation results]{Empirical median and $2.5^{th}$ and $97.5^{th}$ percentiles calculated from the estimates of the CDEs across the 500 simulation datasets using our proposed BKMR-CMA and BKMR-CMA-VS approaches, the linear method, and the traditional method under each simulation scenario. The CDEs presented are for when the mediator is fixed to its 25$^{th}$, 50$^{th}$, and 75$^{th}$ percentiles in the true underlying dataset for each scenario. The truth for each mediation effect and scenario are depicted as black dots. The specific data generation functions used for each simulation scenario are defined in Table 1 of the main text. Results are show for six different data generation scenarios and when the number of mixture components is three and ten.} 
 \label{P2simresults_CDE}
\end{figure}

The empirical median and $2.5^{th}$ and $97.5^{th}$ percentiles across the 500 simulated datasets for the six scenarios are displayed in Figure \ref{P2simresults} for the TE, NDE and NIE and Figure \ref{P2simresults_CDE} for the CDEs. We noticed similar patterns in the empirical distributions for the TE, NDE, and NIE when $L=3$ and $L=10$. We found BKMR-CMA and BKMR-CMA-VS were able to correctly estimate null effects for the NIE and NDE under different causal structures. For scenario 1, we observed equivalent performance of the methods across the mediation effects, as we expected with both $h_M(\cdot)$ and $h_Y(\cdot)$ being linear.  As the scenarios depart from linearity, we observed large differences in the methods across the mediation effects. BKMR-CMA and BKMR-CMA-VS consistently had equal or notably smaller bias than the linear and traditional methods for scenarios 2 through 6. Notably, the $2.5^{th}$ and $97.5^{th}$ percentiles of the linear and traditional approach estimates for the TE under scenarios 3 through 6 did not contain the truth, for the NDE under scenario 3 did not contain the truth, and for the NIE under scenarios 4 through 6 did not contain the truth. When $L=10$, BKMR-CMA-VS was the least biased of all the methods. We observed similar results for the CDEs when we intervene to fix the mediator at its 25$^{th}$, 50$^{th}$, and 75$^{th}$ percentiles in the true underlying datasets as we do for the NDE, with larger differences between BKMR-CMA or BKMR-CMA-VS compared to the linear or traditional methods when the mediator is fixed at higher percentiles. 

\begin{figure}[h]
\hspace*{9mm}\includegraphics[height=2in]{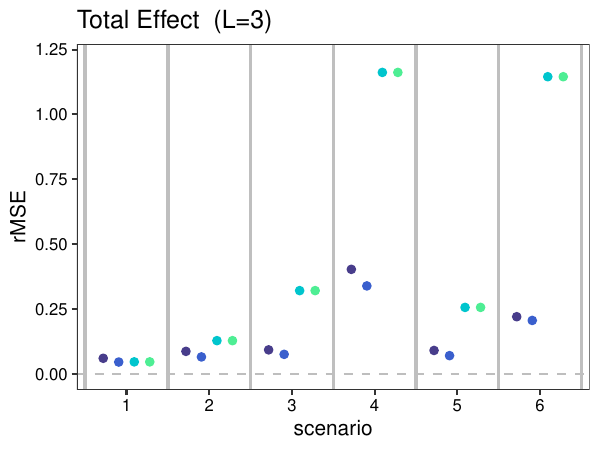}
\includegraphics[height=2in]{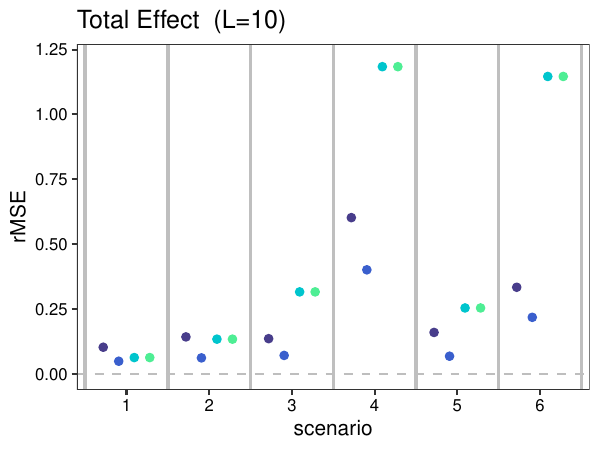}\\

\vspace*{-2mm}\hspace*{9mm}\includegraphics[height=2in]{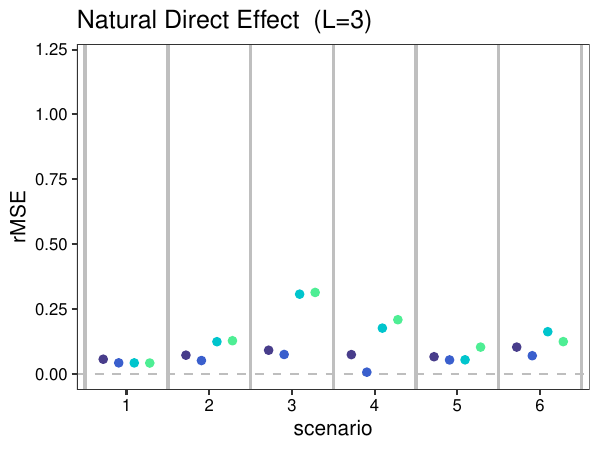}
\includegraphics[height=2in]{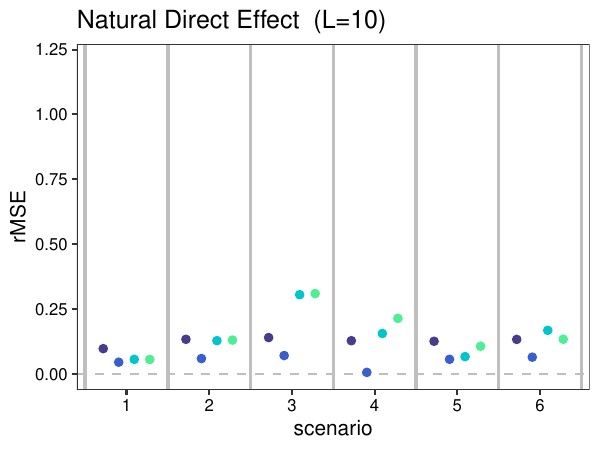}\\

\vspace*{-2mm}\hspace*{9mm}\includegraphics[height=2in]{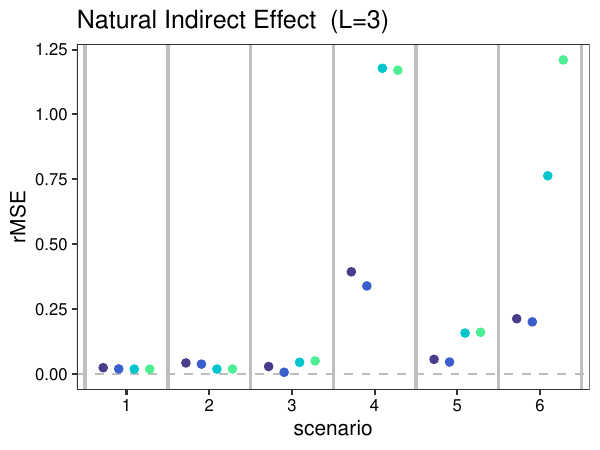}
\includegraphics[height=2in]{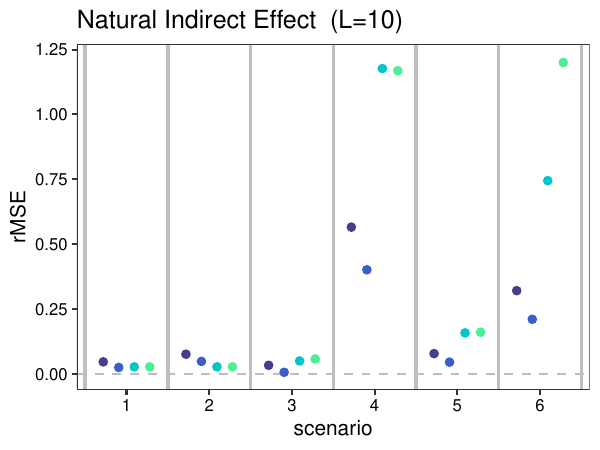}
\includegraphics[height=2in]{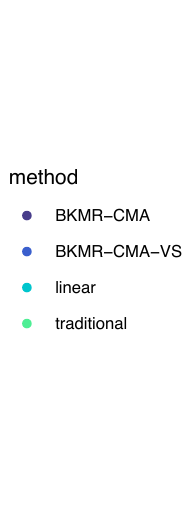}

 \caption[Simulation rMSE values]{ A comparison of the root mean square error (rMSE) from our simulation when the TE, NDE and NIE are estimated by our BKMR-CMA and BKMR-CMA-VS approaches, the linear method, and the traditional method. Results are shown for six different data generation scenarios and when the number of mixture components is three and ten. The specific data generation functions used for each simulation scenario are defined in Table \ref{simscenarios}.} 
 \label{P2simresults_rMSE}
\end{figure}

\begin{figure}[h]
\hspace*{9mm}\includegraphics[height=2in]{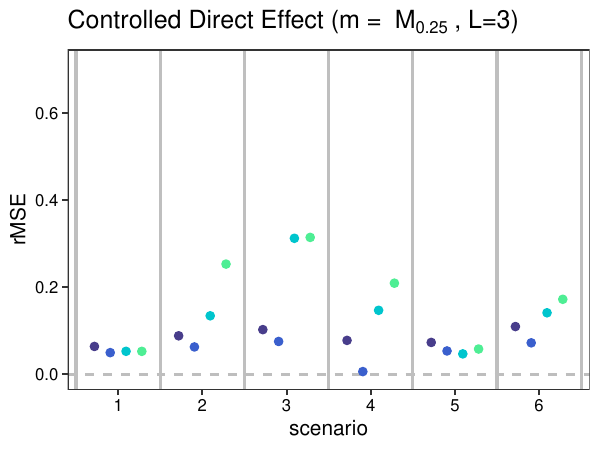}
\includegraphics[height=2in]{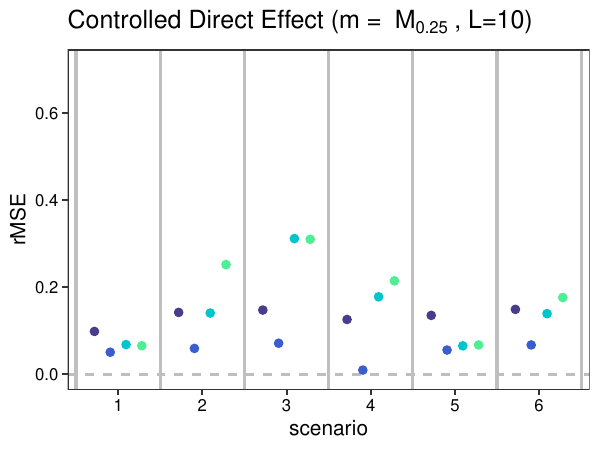}\\

\vspace*{-2mm}\hspace*{9mm}\includegraphics[height=2in]{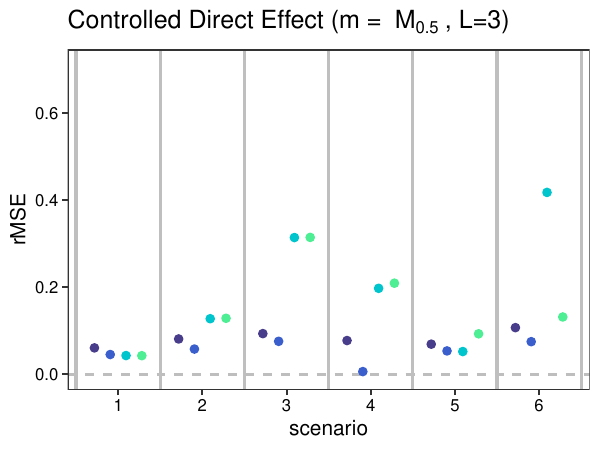}
\includegraphics[height=2in]{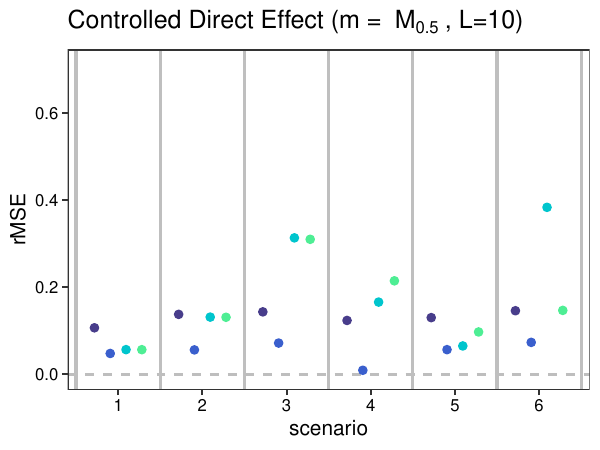}\\

\vspace*{-2mm}\hspace*{9mm}\includegraphics[height=2in]{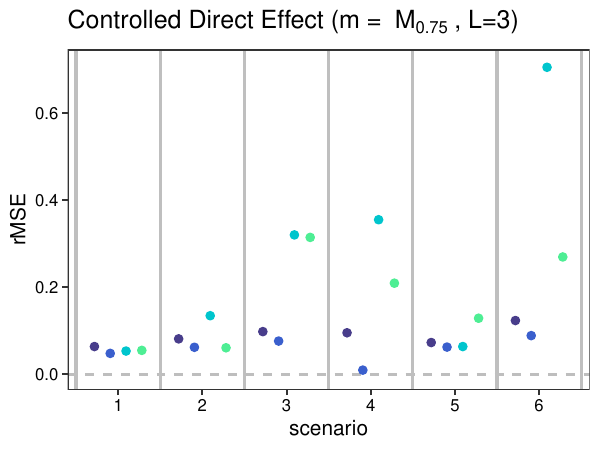}
\includegraphics[height=2in]{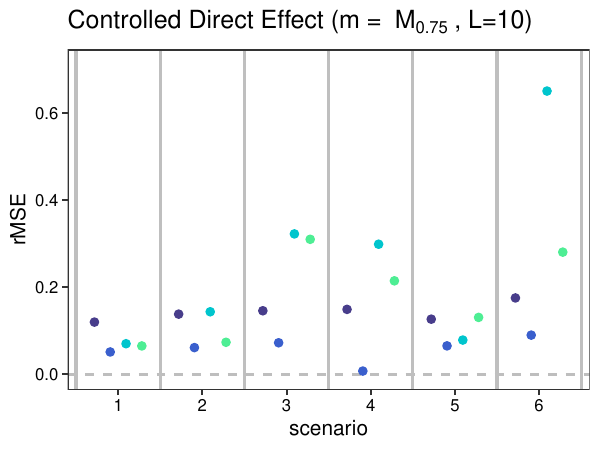}
\includegraphics[height=2in]{P2simlegend_rMSE}

 \caption[Simulation rMSE values]{A comparison of the root mean square error (rMSE) from our simulation when the CDEs are estimated by our BKMR-CMA and BKMR-CMA-VS approaches, the linear method, and the traditional method. The CDEs presented are for when the mediator is fixed to its 25$^{th}$, 50$^{th}$, and 75$^{th}$ percentiles in the true underlying dataset for each scenario. Results are show for six different data generation scenarios and when the number of mixture components is three and ten. The specific data generation functions used for each simulation scenario are defined in Table 1 of the main text.} 
 \label{P2simresults_rMSE_CDE}
\end{figure}

The root mean squared error (rMSE) under the six data generating scenarios we considered are shown in Figure \ref{P2simresults_rMSE} for the TE, NDE and NIE and Figure \ref{P2simresults_rMSE_CDE} for the CDEs. Universally, we observed our BKMR-CMA and BKMR-CMA-VS approaches performed on par or better than current methods in terms of rMSE, especially when the true effect is not null. With a larger number of elements in the mixture, BKMR-CMA-VS performed better than BKMR-CMA. This suggests that as the number of exposures increases and the dose-response surfaces become increasingly more complex, BKMR-CMA and BKMR-CMA-VS outperforms other approaches with respect to estimation within a causal mediation framework. 

\begin{figure}
\hspace*{9mm}\includegraphics[height=2in]{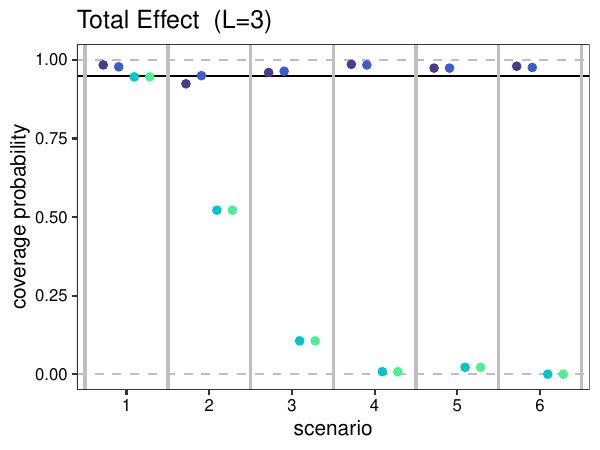}
\includegraphics[height=2in]{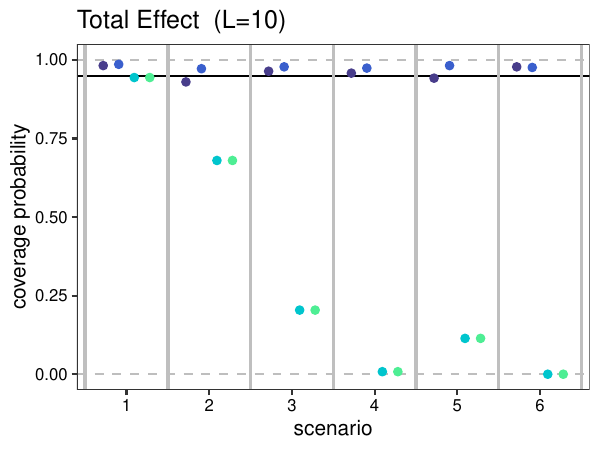}\\

\vspace*{-2mm}\hspace*{9mm}\includegraphics[height=2in]{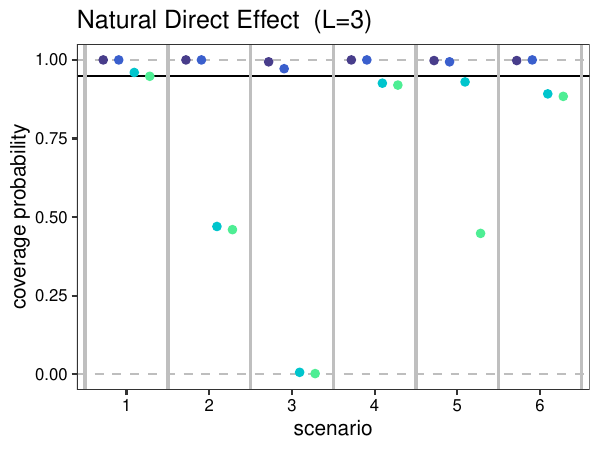}
\includegraphics[height=2in]{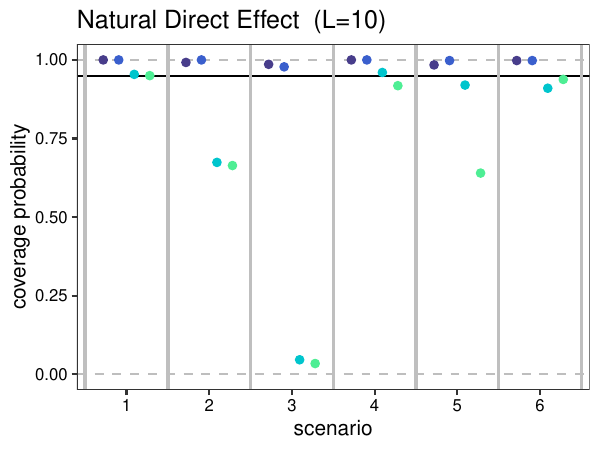}\\

\vspace*{-2mm}\hspace*{9mm}\includegraphics[height=2in]{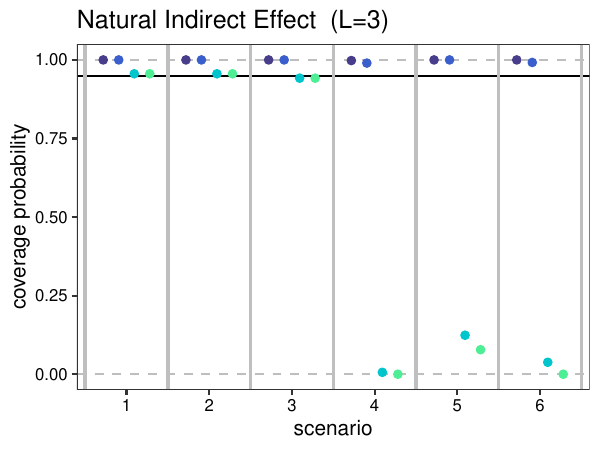}
\includegraphics[height=2in]{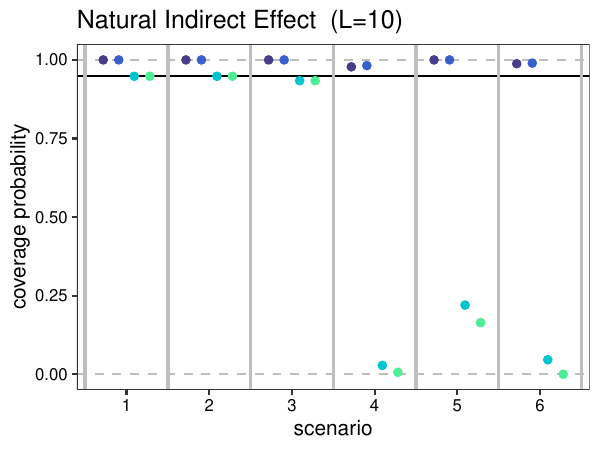}
\includegraphics[height=2in]{P2simlegend_rMSE}

 \caption[Simulation coverage probability]{A comparison of the coverage probability from our simulation when the TE, NDE and NIE are estimated by our BKMR-CMA and BKMR-CMA-VS approaches, the linear method, and the traditional method. The coverage probability is defined as the proportion of the estimates and 95\% credible or confidence intervals in the 500 simulation datasets that contain the truth for each effect. The black line represents a coverage probability of 0.95. Results are shown for six different data generation scenarios and when the number of mixture components is three and ten. The specific data generation functions used for each simulation scenario are defined in Table \ref{simscenarios}.} 
 \label{P2simresults_coverage}
\end{figure}

\begin{figure}[h]
\hspace*{9mm}\includegraphics[height=2in]{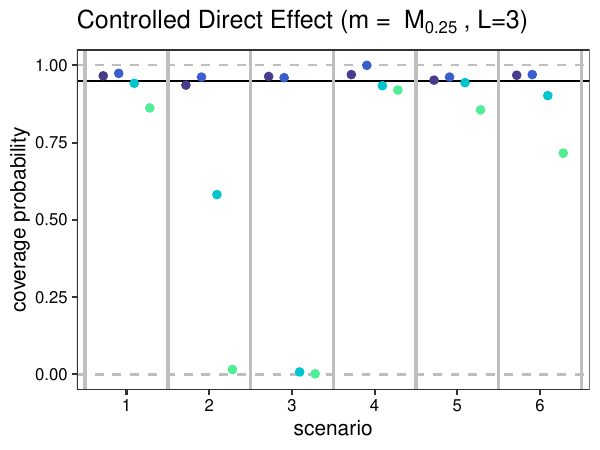}
\includegraphics[height=2in]{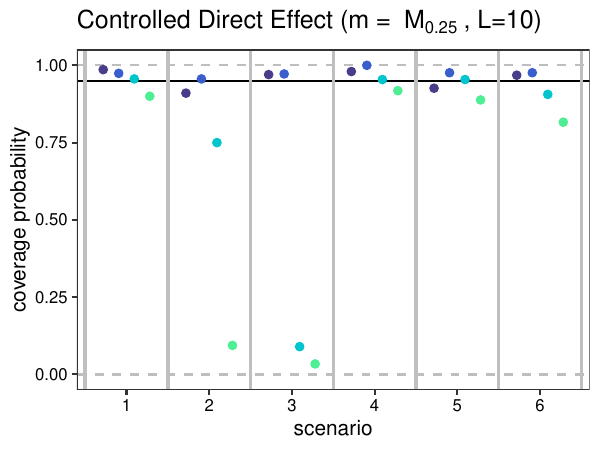}\\

\vspace*{-2mm}\hspace*{9mm}\includegraphics[height=2in]{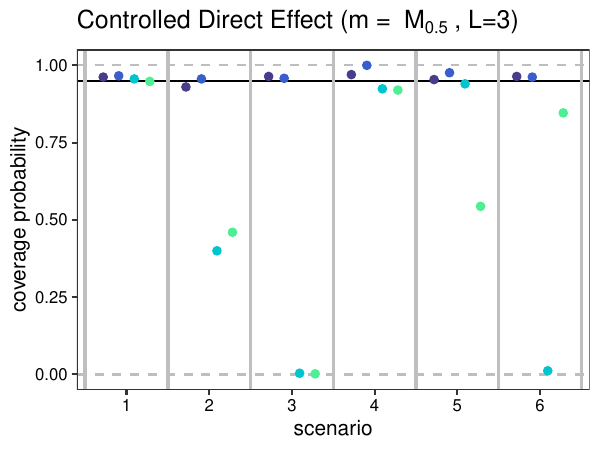}
\includegraphics[height=2in]{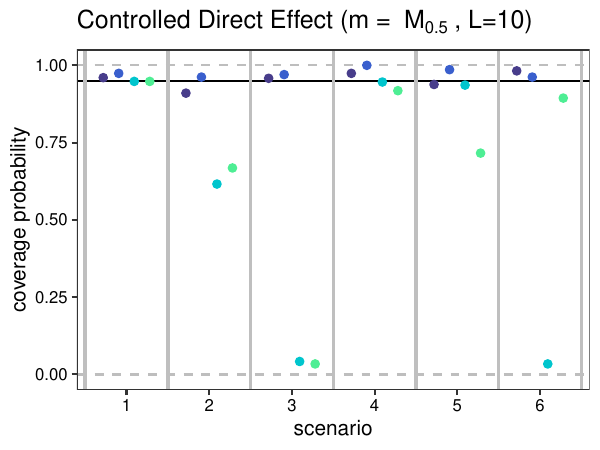}\\

\vspace*{-2mm}\hspace*{9mm}\includegraphics[height=2in]{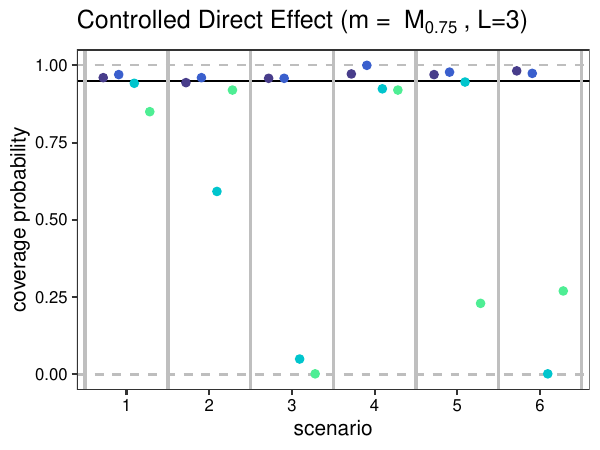}
\includegraphics[height=2in]{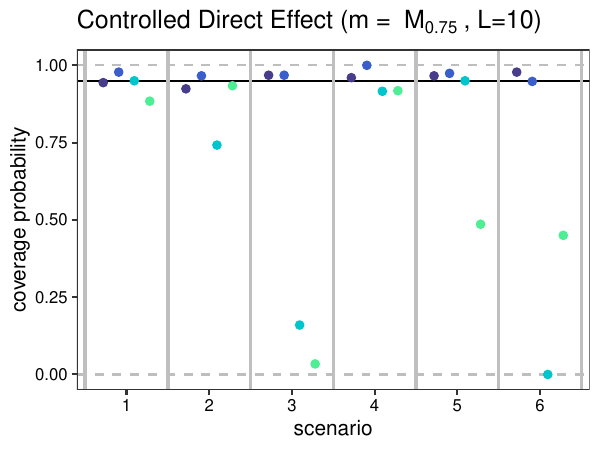}
\includegraphics[height=2in]{P2simlegend_rMSE}

 \caption[Simulation coverage values]{A comparison of the coverage probability from our simulation when the CDEs are estimated by our BKMR-CMA and BKMR-CMA-VS approaches, the linear method, and the traditional method. The coverage probability is defined as the proportion of the estimates and 95\% credible or confidence intervals in the 500 simulation datasets that contain the truth for each effect. The black line represents a coverage probability of 0.95. The CDEs presented are for when the mediator is fixed to its 25$^{th}$, 50$^{th}$, and 75$^{th}$ percentiles in the true underlying dataset for each scenario. Results are show for six different data generation scenarios and when the number of mixture components is three and ten. The specific data generation functions used for each simulation scenario are defined in Table 1 of the main text.} 
 \label{P2simresults_coverage_CDE}
\end{figure}

The coverage probability under the six data generating scenarios are summarized in Figure \ref{P2simresults_coverage} for the TE, NDE and NIE and Figure \ref{P2simresults_coverage_CDE} for the CDEs. For the NDE and NIE, we observed higher than a 0.95 coverage probability of the truth for the $95\%$ credible intervals estimated using BKMR-CMA and BKMR-CMA-VS. Of note, when nonlinear functions were used for $h_M(\cdot)$ and/or $h_Y(\cdot)$ or the true mediation effect was not null, the linear and traditional methods have very low coverage, less than 25\% for many effects and scenarios.


\section{Data analysis}
\label{s:P2data}

\subsection{Study population}
We applied our BKMR-CMA methodology to quantify the contribution of birth length (BL) as a mediator between \emph{in utero} co-exposure to arsenic (As), manganese (Mn) and lead (Pb), and children's neurodevelopment, in a prospective birth cohort in rural Bangladesh. 
This cohort has previously been described.\citep{gleason2014,kile2014,valeri2017} We excluded two mother-infant pairs where the infant had outlying birth lengths (BL $>$ 6 standard deviations from the mean), for a total sample of $727$. Researchers measured \emph{in utero} metal exposure to As, Mn and Pb from umbilical cord venous blood samples. Collaborators in Bangladesh administered the Bayley Scales of Infant and Toddler Development\textsuperscript{TM}, Third Edition (BSID-III\textsuperscript{TM}) to children 20-40 months after birth and neurodevelopment was measured as the raw cognitive development score (CS).\citep{bayley2006} We controlled for clinic, child sex, maternal IQ, maternal education (less than high school vs. at least high school), maternal protein intake (low vs. medium vs. high tertiles), secondhand smoke exposure at baseline (smoking environment vs. non-smoking environment), HOME score, and maternal age at delivery in all analyses. When conducting our analyses, we log transformed and standardized metal concentrations, and standardized CS, BL, and other continuous variables. To evaluate the persistence of the effect of \emph{in utero} metal exposure on neurodevelopment, we considered age at BSID-III testing to be an effect modifier of the metal mixture and neurodevelopment relationship. 

\subsection{Models}
We modeled the effect of co-exposure to As, Mn and Pb on birth length via BKMR mediator model (\ref{P2:medmodel}), where $\bE_M$ did not include any variables. To model the joint effect of the metal mixture and birth length on neurodevelopment, we fit a BKMR outcome model (\ref{P2:outmodel}) with all three metals, age at BSID-III testing and birth length in the kernel function ($\bE_Y$ only included age at BSID-III testing). We examined the relationship between the metal mixture on birth length and the relationship between the metal mixture, child's age and birth length on neurodevelopment. 

We simulated counterfactuals to estimate the NDE, NIE, and TE for a change in the raw (untransformed) exposures from $\ba^*=(As_{.25}=0.56 \mu$g/dL, $Mn_{.25}= 4.72 \mu$g/dL, $Pb_{.25}= 1.15 \mu$g/dL), all metals set at their corresponding 25$^{th}$ percentile, to $\ba=(As_{.75}= 1.58 \mu$g/dL, $Mn_{.75}= 17.80 \mu$g/dL, $Pb_{.75}=2.42 \mu$g/dL), all metals set at their 75$^{th}$ percentile, when age at testing was set at its 10$^{th}$ percentile of 24.6 months and its 90$^{th}$ percentile of 31.0 months. We also calculated the CDEs for a change in exposure from $\ba^*$ to $\ba$ when birth length was set to its 10$^{th}$ percentile of 44cm, median value of 46cm, and 75$^{th}$ percentile of 48cm, fixing age at testing to 24.6 and 31.0 months. We used three dimensional scatterplots to verify there are observations in the multivariate exposures space to justify these values of $\ba^*$ and $\ba$ for our data application.

We estimated the mediation effects using both BKMR-CMA and BKMR-CMA-VS and compared these results to the effects estimated by (1) the linear method including an age at testing by Mn interaction (motivated by our BKMR analysis), (2) the linear method without an age at testing by Mn interaction, and (3) the traditional method. For methods 1-3 listed above, we included child's age at the time of the Bayley Scale administration and a child's age squared when modeling children's neurodevelopment.


\subsection{Results}
\begin{figure}[h]
  \centering
  \includegraphics[width=4in]{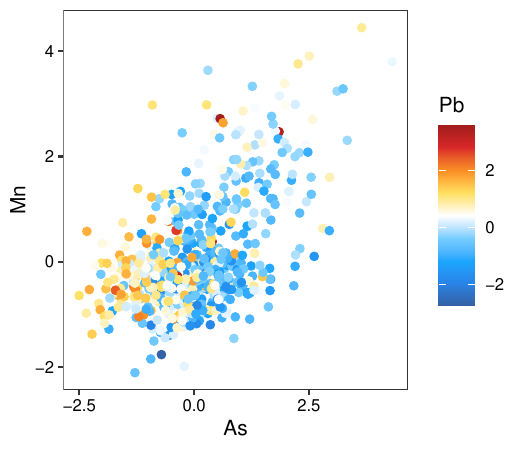}
 \caption{Three dimensional scatter plot of natural logarithm transformed and standardized metal concentrations observed in our Bangladeshi cohort. The 25$^{th}$ percentiles of the natural logarithm transformed and standardized metals are $As_{.25}=-0.61,$ $Mn_{.25}= -0.64,$ $Pb_{.25}= -0.80$, which corresponds to the raw levels of $As_{.25}=0.56 \mu$g/dL, $Mn_{.25}= 4.72 \mu$g/dL, $Pb_{.25}= 1.15 \mu$g/dL.  The 75$^{th}$ percentiles of the natural logarithm transformed and standardized of metals are $As_{.75}=0.55,$ $Mn_{.75}= 0.34,$ $Pb_{.75}= 0.74$, which corresponds to the raw levels of $As_{.75}= 1.58 \mu$g/dL, $Mn_{.75}= 17.80 \mu$g/dL, $Pb_{.75}=2.42 \mu$g/dL. } 
 \label{3Dplot}
\end{figure}

\begin{figure}[h]

\hspace*{3mm}\includegraphics[height=2.9in]{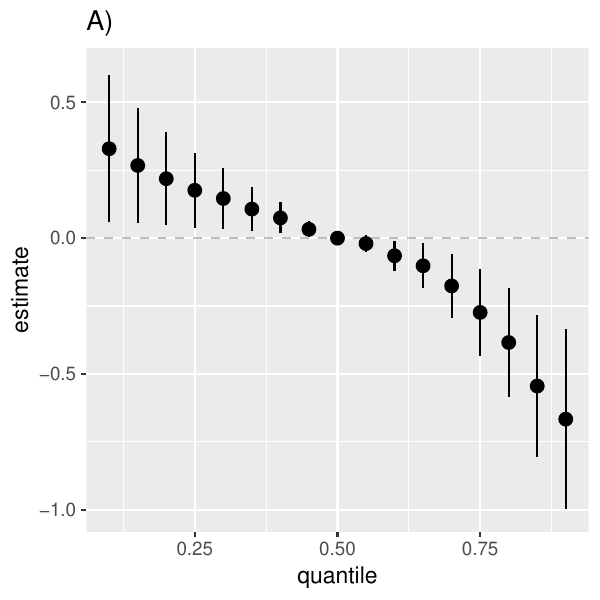}
\includegraphics[height=2.9in]{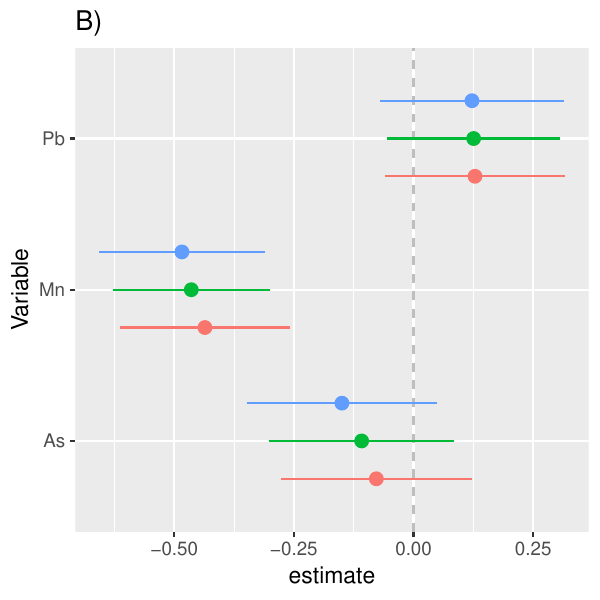}\\

\vspace{-6mm}\hspace*{3mm}\includegraphics[height=2.9in]{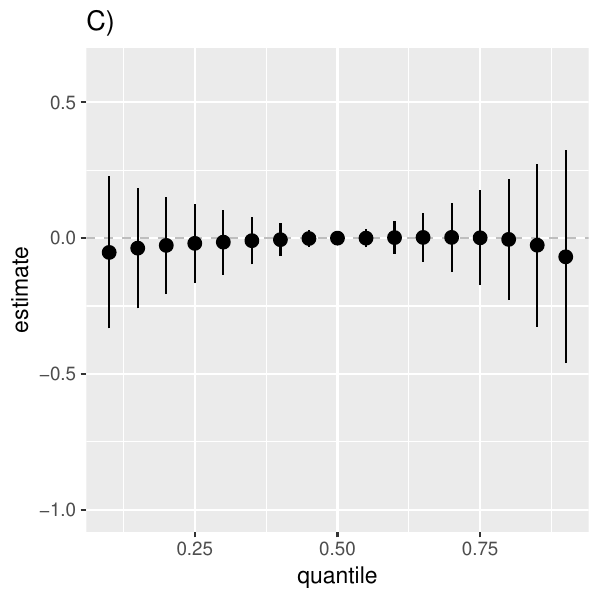}
\includegraphics[height=2.9in]{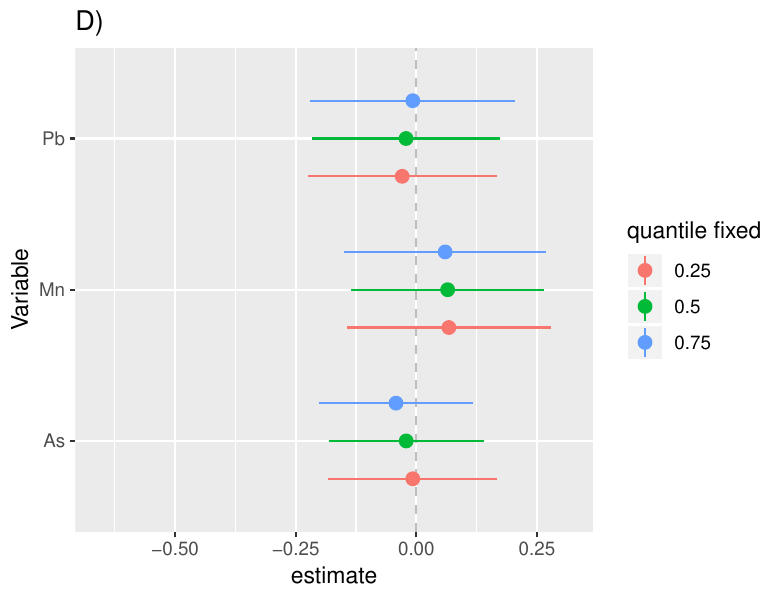}\\

 \caption[]{ The joint effect of co-exposure to As, Mn and Pb on neurodevelopment and single variable effects observed in our Bangladeshi cohort estimated by BKMR. Presented when age is fixed at its 10$^{th}$ percentile of 24.6 months (A,B) and $90^{th}$ percentile of 31.0 months (C,D). The average change in cognitive score for a joint change in the metal mixture from the quantile value on the $x$-axis to the median (estimates and 95\% CI), when age is fixed at 24.6 months (A) and 31.0 months (C). Single metal associations with neurodevelopment (estimates and 95\% CI, gray dashed line a the null). These figures show the average change in cognitive score for a change in a single metal from its 25$^{th}$ to 75$^{th}$ percentile, fixing the other metals at their 25$^{th}$, 50$^{th}$, or 75$^{th}$ percentiles, when age is fixed at 24.6 months (B) and 31.0 months (D).}
 \label{P2_TE_plots}
\end{figure}

Figure \ref{3Dplot} is a three dimensional scatterplot showing there are children who had low levels of all three metals as well as children who had high levels of all three metals in our Bangladeshi cohort. This figure shows that our contrast of interest lies within the multidimensional support of our data. Figure \ref{P2_TE_plots} shows age at testing modifies the relationship between the metal mixture and neurodevelopment. We observed a significant harmful effect of the metal mixture on neurodevelopment when age was fixed at its 10$^{th}$ percentile value of 24.6 months and a null effect when age was fixed to its 90$^{th}$ percentile value of 31.0 months. The significant harmful effect seen when age of testing was fixed at 24.6 months was driven by Mn (Figure \ref{P2_TE_plots}B).  

\begin{figure}[h]
\centering

\includegraphics[height=4.5in]{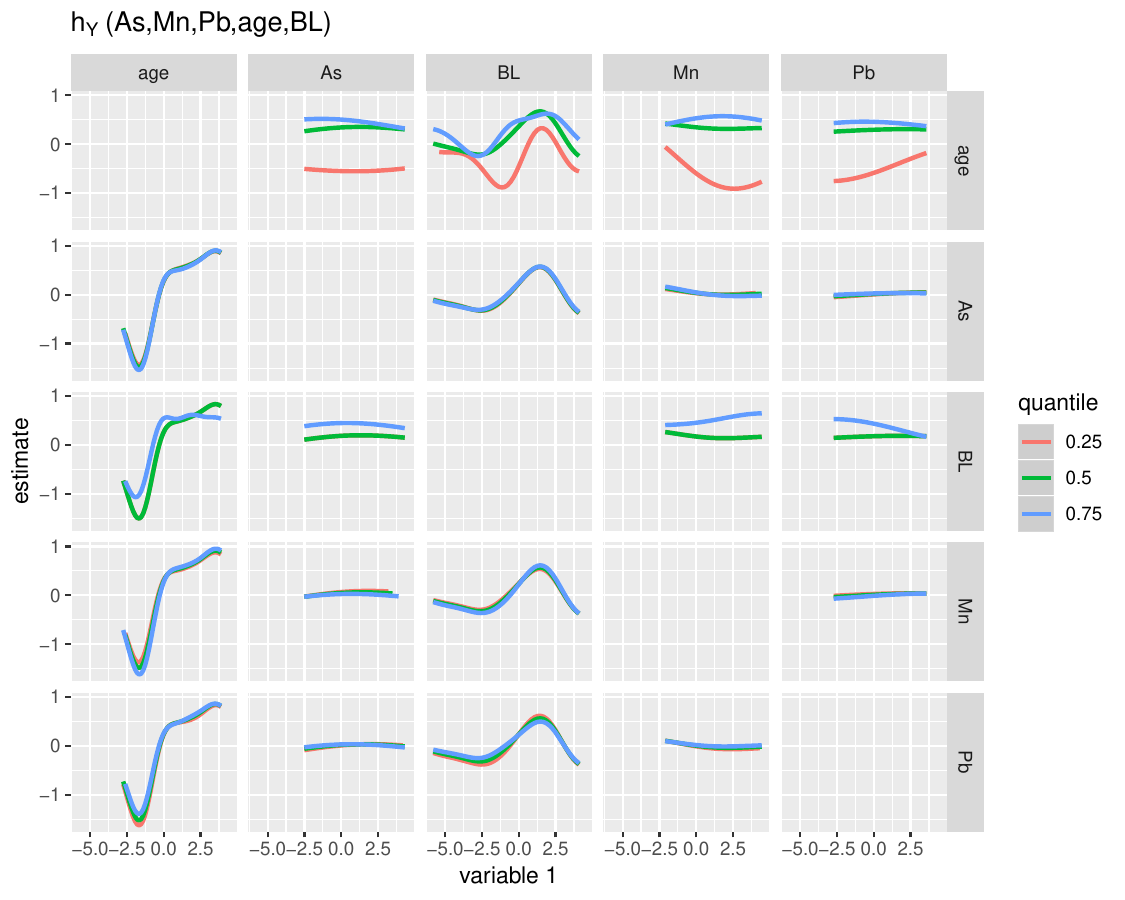}\\

 \caption[]{ Bivariate associations between each column element (listed on the top) and neurodevelopment, when the row element (listed on the right) is fixed at its 25$^{th}$, 50$^{th}$ or $75^{th}$ percentiles, and the remaining elements are fixed at their median. } 
 \label{P2_bivar_y}
\end{figure}

Figure \ref{P2_bivar_y} displays the bivariate effect on neurodevelopment of each element listed on the top when the element listed on the right was fixed at its 25$^{th}$, 50$^{th}$ or $75^{th}$ percentiles, and all remaining elements were fixed at their median. We observed nonlinear associations of age at testing and BL with neurodevelopment and interactions between age and both birth length and Mn. There also appeared to be small interactions of Mn and Pb with birth length on neurodevelopment. 

\begin{figure}[h]

\hspace*{5mm}\includegraphics[height=2.6in]{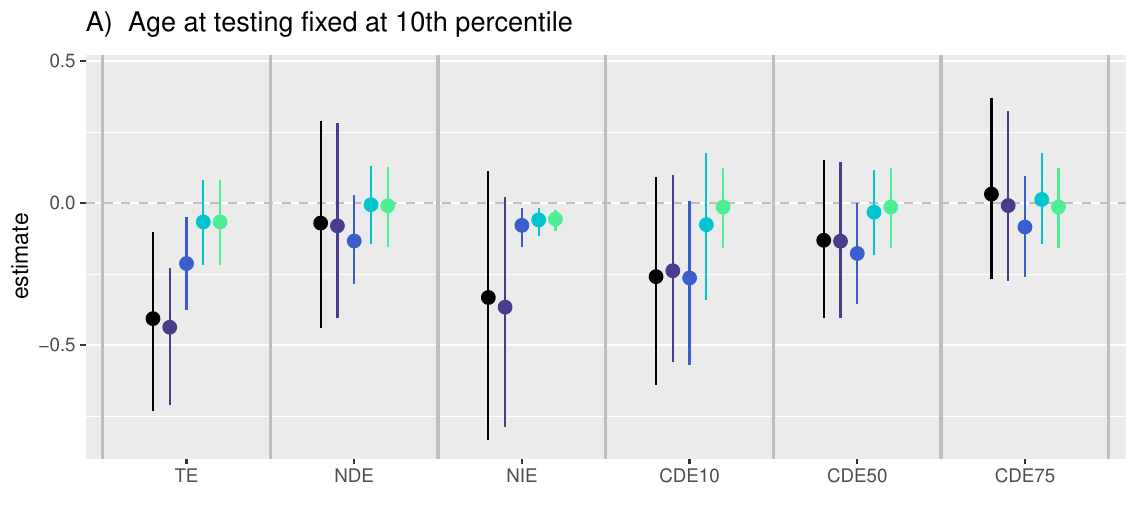}\\

\vspace*{-6mm}\hspace*{5mm}\includegraphics[height=2.6in]{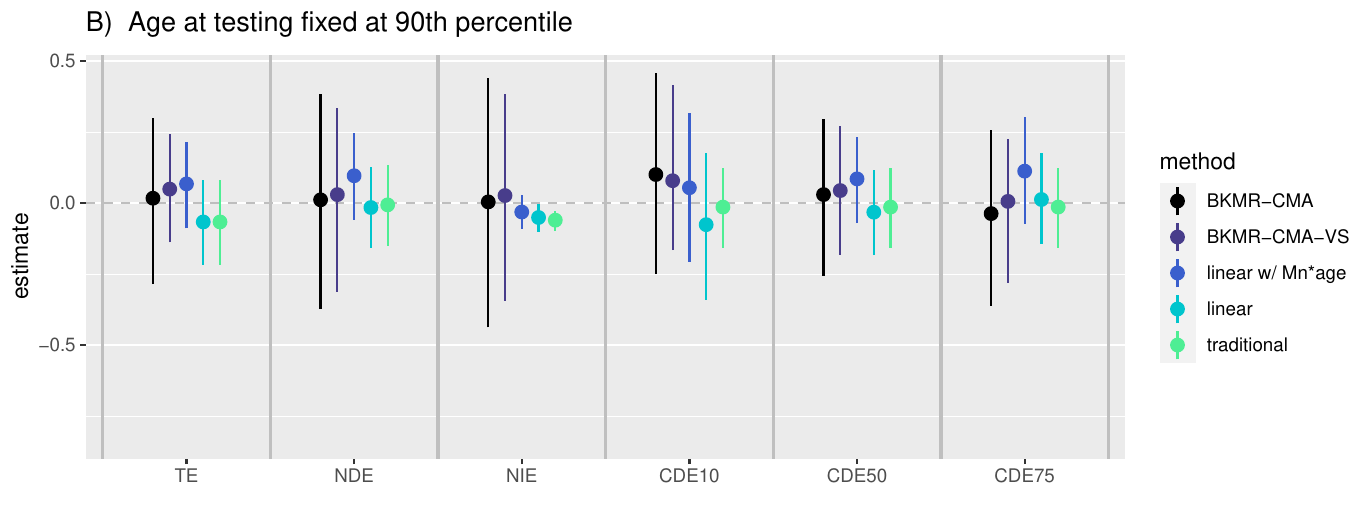}\\

 \caption[Mediation effects estimated in our Bangladeshi cohort using BKMR-CMA.]{Mediation effects estimated in our Bangladeshi cohort using BKMR-CMA, BKMR-CMA-VS, the linear method with and without an age by Mn interaction, and the traditional method. All effects are estimated for a change of the metal mixture from its raw (untransformed) $25^{th}$ percentile $\ba^*=(As_{.25}=0.56 \mu$g/dL, $Mn_{.25}= 4.72 \mu$g/dL, $Pb_{.25}= 1.15 \mu$g/dL) to its raw $75^{th}$ percentile $\ba=(As_{.75}= 1.58 \mu$g/dL, $Mn_{.75}= 17.80 \mu$g/dL, $Pb_{.75}=2.42 \mu$g/dL) and fixing age at its 10$^{th}$ percentile of 24.6 months (A) and $90^{th}$ percentile of 31.0 months (B). The CDEs are calculated as the direct effect from $\ba^*$ to $\ba$ intervening to fix the mediator at its $10^{th}$, $50^{th}$, $75^{th}$ percentiles values of 44, 46, and 48cm respectively.} 
 \label{P2_mediationeffects}
\end{figure}

Figure \ref{P2_mediationeffects} summarizes the mediation effects for a change of the raw metal mixture from $\ba^*=(As_{.25}=0.56 \mu$g/dL, $Mn_{.25}= 4.72 \mu$g/dL, $Pb_{.25}= 1.15 \mu$g/dL) to $\ba=(As_{.75}= 1.58 \mu$g/dL, $Mn_{.75}= 17.80 \mu$g/dL, $Pb_{.75}=2.42 \mu$g/dL) when age at testing was either fixed at 24.6 or 31.0 months. We observed substantial differences between the results for the two age at testing percentiles when the mediation effects were estimated with more flexible methods. Specifically, we observed a large TE and NIE for both our BKMR-CMA and BKMR-CMA-VS approaches when age at testing was fixed at 24.6 months (Figure \ref{P2_mediationeffects}A), however, this effect was null when age at testing was fixed to 31 months (Figure \ref{P2_mediationeffects}B). 

When age of testing was fixed to 24.6 months, BKMR-CMA estimated the TE of an interquartile range (IQR) change in the metal mixture to correspond to a 0.41 standard deviation decrease in cognitive scores (95\% CI = [-0.73, -0.10]). The NDE, the causal effect of the IQR change of the metal mixture on cognitive score had birth length been set to the level it would take if the metal mixture was at its 25$^{th}$ percentile, that is the effect that operates independently from birth length, was estimated to be -0.07 (95\% CI = [-0.44, 0.29]). Additionally, the NIE, the causal effect of a change in birth length induced by a IQR change in the metal mixture and had the metal mixture exposure been set to its 75$^{th}$ percentile (i.e., the effect that operates indirectly through or mediated by birth length) was estimated to be -0.33 (95\% CI = [-0.83, 0.11]). Thus, we found that the majority of the adverse effect of the metal mixture on cognitive function operates through birth length when age at testing was fixed to 24.6 months (percent mediated $=81.7\%$). \\

The CDEs, the direct effect of an IQR change of the metal mixture under a hypothetical intervention that fixes birth length for everyone at specific quantiles of that distribution, showed that among children who were tested earlier in life, the harmful direct effect of the metal mixture on neurodevelopment was reduced for higher quantiles of birth length. For instance, had we fixed birth length for everyone to the 10$^{th}$ percentile value of 44cm,  we estimated the direct effect of an IQR change of the mixture on cognitive function to be -0.26 (95\% CI = [-0.64, 0.09]). However, if we had fixed birth length for everyone to the 75$^{th}$ percentile value of 48cm,  we estimated the direct effect of an IQR change of the mixture on cognitive function to be null ($CDE_{M_{0.75}}=0.03$, 95\% CI = [-0.27, 0.15]). This suggests that intervention on fetal growth could potentially reduce the harmful impact of this metal mixture on children's neurodevelopment.


\section{Discussion}
\label{P2:discuss}

We have proposed BKMR-CMA as a way to estimate the direct and indirect effects of an environmental mixture on an outcome through an intermediate variable. To our knowledge, this is the first method presented in the causal inference literature to estimate these effects when the exposure of interest is a potentially complex mixture, without \emph{a priori} knowledge of the exposure-mediator and exposure-mediator-outcome relationships. This method allows for complex relationships between the elements of the mixture and the mediator, and between mixture elements, the mediator, continuous effect modifiers, and the outcome through a joint kernel specification, and inherently controls for multiplicity. The nonparametric treatment of these relationships avoids explicit specification of parametric terms for all of the potential nonlinear and/or interaction terms in the high-dimensional exposure-response function. 
Our extension of causal mediation methodology that allows for a mixture of exposures is important for many environmental health applications. In the Bangladesh application, the use of our proposed BKMR-CMA approach uncovered patterns in the data that would have been completely missed using simpler, existing methods. 

We estimate the TE, NDE, NIE, and CDEs at fixed levels of the modifiers and for a fixed intervention in the exposures through simulation of counterfactuals from the posterior predictive distribution for each MCMC iteration and make inference from these posterior samples of the mediation effects. Our simulations show that our proposed BKMR-CMA and BKMR-CMA-VS approaches outperform current methods when the underlying data generation mechanisms are not adequately modeled as a collection of linear models. Additionally, our simulations show our method is unbiased even when complex relationships do not exist. As with any semiparametric formulation, when the true data generating mechanism is closely approximated by a simpler parametric model, it will not be as efficient as the parametric model. In the presence of complex data generation scenarios, we advise to use our approach over other methods, as BKMR-CMA does not place assumptions on the relationships between the mixture elements, mediator and outcome. In this work our goal was estimation, so we focused on the performance of BKMR-CMA and BKMR-CMA-VS with respect to bias, rMSE, and interval coverage.  If one were explicitly interested in hypothesis testing, other frequentist approaches may be preferable.  We do note that Scott and Berger (2010) explained that treating the probability parameter in the spike-and-slab formulation of the variable selection procedure employed by BKMR ($\pi$ in Bobb et al. 2015) as an unknown parameter estimated from the data yields an automatic multiple-testing penalty.\citep{scott2010, bobb2015}

Applying these methods to a prospective Bangladeshi birth cohort, we found a negative (adverse) association of co-exposure to lead, arsenic, and manganese on neurodevelopment, a negative association of exposure to this metal mixture on birth length, evidence that birth length mediates the effect of this metal mixture on children's neurodevelopment, and that age modifies these relationships. If birth length were fixed to its $75^{th}$ percentile value of 48cm and age fixed to its 10$^{th}$ percentile value of 24.6 months, the harmful direct effect of the metal mixture on neurodevelopment is attenuated compared to smaller birth length percentiles, suggesting that targeted interventions to promote healthy fetal growth could potentially block part of the adverse effect of metals on neurodevelopment. 
For those children exhibiting effects (those at the 10th percentile of age at testing), the BKMR-CMA and BKMR-CMA-VS analyses suggests a relatively large total estimated effect of the mixture, and that this effect is largely due to an indirect, rather than a direct, effect. Estimates from existing approaches were severely attenuated towards the null, even when an age squared term and age by Mn interaction is included in the model for cognitive score, which, while they might have confidence intervals that do not contain zero, presented an entirely different picture of the impact of this metal mixture on neurodevelopment in this Bangladeshi cohort. This demonstrates the importance of using a flexible method, such as our BKMR-CMA or BKMR-CMA-VS approach, when estimating mediation effects of a mixture.

Prior to implementing BKMR-CMA or BKMR-CMA-VS, careful thought should be given via subject matter expertise and a literature review to determine the mixture set and potential effect modifiers to include in $h_M(\cdot)$ and $h_Y(\cdot)$. As with all statistical analyses (mediation and otherwise), when the exposure of interest is continuous, there are infinite options to consider for the scale on which the effects are reported. It is best to define the contrast of interest through extensive data exploration and visualization of the multidimensional exposure distribution \emph{a priori} to data analysis to avoid intentionally selecting the contrast that reports the largest effect. It is important that there is enough data to evaluate the hypothetical intervention of interest without model extrapolation\citep{zigler2021}. As it is common in the field of environmental epidemiology, an interquartile range change in exposure could be chosen for $\ba^*$ and $\ba$. In our experience, using the 25$^{th}$ and 75$^{th}$ (or 25$^{th}$ and 50$^{th}$) percentiles work well for these types of environmental mixtures since they typically exhibit moderate to high correlations between the mixture elements. With moderate to high correlation among the mixture components, the exposure concentrations are more likely to be similar. Thus, we would expect some observations where all of the exposures levels are low and some where all are high. When some components of the mixture are weakly correlated, we recommend the investigator focuses on understanding the distributions of these weakly correlated mixture components through extensive data visualization and then chooses a contrast that represents differences between groups of subjects in the study population. Of note, for effects that appear quadratic, reporting effects per the interquartile range has the potential to mask effects of a given exposure on an outcome. In such cases, one could consider contrasts from the $10^{th}$ or $25^{th}$ percentiles to the median. It is important to assess whether the final contrast chosen is in the support of the observed data to avoid extrapolation of the models. 
	
BKMR-CMA is useful in several settings, including when there is a high dimensional exposure, when there are multiple exposures and effect modifiers, or both. One can use the hierarchical variable selection option in the $\tt{bkmr}$ package in addition to component-wise variable selection presented here. In addition to effect modifiers, our general framework can allow for complex relationships between confounders, the exposures, mediator and outcome, by also including confounding variables in the kernel functions. However, in this setting, one would need to integrate over the distribution of the confounders included in the kernel function to obtain correct marginal estimates (instead of evaluating the effects at the mean level of these confounders). 

A limitation of our method is the increasing computation time required to fit BKMR and predict counterfactuals as the complexity of the model and sample size increase. The computational burden of the BKMR-CMA algorithm for a single simulation dataset with $N=300$ is just over four hours.

In some applications, exposure to mixtures with more than ten elements is common. When the number of mixture elements is significantly greater than 10, simulations studies would need be conducted to see how our method performs. In the current formulation of our algorithm, we assume the mediator and outcome models are independent. This is a common assumption in causal mediation literature, however, this is a limitation of our methods. Although models (9) and (10) are theoretically compatible, we acknowledge the phenomenon noted by Sanders and Blume (2018) whereby a model may be structurally compatible but the estimates from the conditional and total effect models do not exhibit compatibility (i.e. do not match) likely due to sampling variability in finite samples.\citep{saunders2018} We note this is an issue in all causal mediation models, and not unique to BKMR-CMA or BKMR-CMA-VS. This could also occur if the true exposure-response functions are not within the space of functions spanned by BKMR, which is always a possibility but likely rare due to the smoothness often observed in biological effects of environmental exposures. In our data application, our results are limited by potential residual confounding by malnutrition and measurement error in birth length. In the presence of measurement error in the mediator, our methods would need to incorporate current methods for measurement error in mediation.\citep{valeri2014,valeri2017epig}  

In future work, we plan to consider joint specification of the mediator and outcome models to reduce the assumptions needed for BKMR-CMA to be interpreted causally. Additionally, we hope to extend BKMR-CMA to allow for discrete and failure time mediator and outcome distributions as well as to allow for multiple mediators and/or multiple outcomes. Further, BKMR-CMA could be extended  to consider a stochastic or distributional change on the exposures instead of a fixed intervention. Overall, we find BKMR-CMA to be a potentially useful tool to estimate mediation effects when the exposure of interest is an environmental mixture.


\section*{Acknowledgements}
This work was supported by the National Institutes of Health [grant numbers ES007142, ES028800, ES026555, ES000002, ES023515, ES016454, ES013744, ES28522, ES017437, ES015533, K01MH118477].



\bibliography{BKMR-CMA}
\bibliographystyle{apalike}

\appendix
\numberwithin{equation}{section}


\newpage
\section{Algorithm to estimate CDEs using BKMR-CMA}
\label{P2:CDEalgorithm}

\begin{enumerate}
	\item Fit BKMR outcome model (9).
			
	\item For each MCMC iteration, $j=1,\hdots, J $:
			
	\begin{enumerate}
\item Estimate the average outcome value for the mean level of covariates at the specific mediator value of interest for $\tilde{\bz}=\begin{pmatrix} \bz_Y & \bz_Y^* \end{pmatrix}^T$ from (9) (i.e. estimate $Y_{\bz^*m}$ and $Y_{\bz m}$ for each MCMC iteration).
	\vspace*{-3mm}\begin{eqnarray*}
		\bY_{\tilde{\bz} m}^{(j)}(\cbar) &=& \E^{(j)}(Y | \bZ_Y = \tilde{\bz}, M= m, \bC=\cbar)\\
		& =& h_Y^{(j)}(\bZ_Y =\tilde{\bz},M=m) + \cbar^T\btheta^{(j)}
	\end{eqnarray*}
								
	\vspace*{-3mm}\item Obtain the $j^{th}$ posterior sample of the CDE for a change of exposure from $\ba^*$ to $\ba$ intervening to fix the mediator at $m$ and effect modifiers $\bE_Y$ at $\be_Y$ by:							
	\vspace*{-3mm}\begin{eqnarray*}
				\text{CDE}(m)^{(j)} &=&  Y_{\bz m}^{(j)}(\cbar)- Y_{\bz^*m}^{(j)}(\cbar).
\end{eqnarray*}	
		\end{enumerate}
			
	\item Estimate the CDE$(m)$ and its 95\% credible interval conditional for level of the effect modifiers and marginally according to the confounders as the posterior mean and posterior $2.5^{th}$ and $97.5^{th}$ percentiles from these posterior samples.
\end{enumerate}
\noindent Only two no unmeasured confounding assumptions are required for the CDE to have a causal interpretation: $Y_{\ba m} \amalg \bA | \bC$, $Y_{\ba m} \amalg M | \bC, \bA$. Namely, there are no unmeasured exposure-outcome confounders and there are no unmeasured mediator-outcome confounders.


\section{Formulas to estimate causal mediation effects when the exposure is a mixture}
\label{s:causalmedformulas}

\noindent Consider the following linear regression models for the mediator and outcome:
\vspace*{-3mm}\begin{eqnarray} \label{medmodelLM}
	\E[M] &=& \beta_0 + \bbeta_1^T\bA +\bbeta_{2}^T\bC, \\ \label{outmodelLM}
	\E[Y] &=& \theta_0 + \btheta_1^T\bA + \theta_2M + \btheta_3^T\bA M+  \btheta_4^T\bC, 
\end{eqnarray}

\vspace*{-3mm}\noindent where $\bA=(A_1, \dots, A_L)^T$ is a exposure mixture of $L$ components, $\bbeta_1=(\beta_{11}, \dots \beta_{1L})^T$,  $\btheta_1=(\theta_{11}, \dots \theta_{1L})^T$, and $\btheta_3=(\theta_{31}, \dots \theta_{3L})^T$.\\

\noindent $\E_Y(Y_{\ba m} | \bC=\bc)$ represents the expected outcome value had everyone been exposed to level $\ba$ and had their mediator been set to level $m$, fixing covariates to level $\bc$. $\E_Y(Y_{\ba M_{\ba^*}} | \bC=\bc)$ represents the expected outcome value had everyone been exposed to level $\ba$ and had their mediator been set to the level it would have taken if exposure is set to $\ba^*$, fixing covariates to level $\bc$. Then, considering models (\ref{medmodelLM}) and (\ref{outmodelLM}), and assuming (i) $Y_{\ba m} \amalg \bA | \bC$, (ii) $Y_{\ba m} \amalg M | \bC, \bA$, (iii) $M_{\ba} \amalg \bA | \bC$, and (iv) $Y_{\ba m} \amalg M_{\ba^*} | \bC$, we can estimate these effects as:\\
	
	\vspace*{-1.5cm} \begin{eqnarray*}
		~~\E_Y(Y_{\ba m} | \bC )
		 	&\overset{(i)-(ii)}{=}& \E_Y(Y | \bA=\ba, M=m, \bC = \bc) ~~~~~~~~\text{by consistency}\\
			&=& \theta_0 + \btheta_1^T\ba + \theta_2m + \btheta_3^T\ba m+  \btheta_4^T\bc \\
		~\E_Y(Y_{\ba^*m} | \bC )
		 	&\overset{(i)-(ii)}{=}& \E_Y(Y | \bA=\ba^*, M=m, \bC = \bc) ~~~~~~~\text{by consistency}\\
			&=& \theta_0 + \btheta_1^T\ba^* + \theta_2m + \btheta_3^T\ba^* m+  \btheta_4^T\bc 
	\end{eqnarray*} 

	\vspace*{-1cm} \begin{eqnarray*}
		\E_Y(Y_{\ba M_{\ba^*}} | \bC ) &=& \int_m \E_Y(Y_{\ba m} | M_{\ba^*} = m, \bC = \bc) dP_{M_{\ba^*}}(m|\bC=\bc)\\
			&\overset{(iii)}{=}&\int_m \E_Y(Y_{\ba m} | M_{\ba^*} = m, \bC = \bc) dP_{M_{\ba^*}}(m|\bA = \ba^*, \bC=\bc)\\
			&\overset{(iv)}{=}&\int_m \E_Y(Y_{\ba m} | \bC = \bc) dP_{M}(m|\bA = \ba^*, \bC=\bc) ~~~~~~~~~~~~~~~~~~~~~~\text{by consistency}\\
			&\overset{(i)-(ii)}{=}&\int_m \E_Y(Y | \bA=\ba, M=m, \bC = \bc) dP_{M}(m|\bA = \ba^*, \bC=\bc) ~~~~\text{by consistency}
			\end{eqnarray*} 
			\begin{eqnarray*}
			&=&\int_m \theta_0 + \btheta_1^T\ba + \theta_2m + \btheta_3^T\ba m+  \btheta_4^T\bc 
			~dP_{M}(m|\bA = \ba^*, \bC=\bc) \\
			&=& \theta_0 + \btheta_1^T\ba +  \btheta_4^T\bc + \left(\theta_2 + \btheta_3^T\ba\right)\int_m  m ~dP_{M}(m|\bA = \ba^*, \bC=\bc) \\
			&=& \theta_0 + \btheta_1^T\ba +  \btheta_4^T\bc + \left(\theta_2 + \btheta_3^T\ba\right)\E_M(M |\bA = \ba^*, \bC=\bc) \\
			&=& \theta_0 + \btheta_1^T\ba +  \btheta_4^T\bc + \left(\theta_2 + \btheta_3^T\ba\right) \left[\beta_0 + \bbeta_1^T\ba^* +\bbeta_{2}^T\bc\right] 
	\end{eqnarray*} 

By similar logic, 
\begin{eqnarray*}
\E_Y(Y_{\ba M_{\ba}} | \bC )      &=& \theta_0 + \btheta_1^T\ba    +  \btheta_4^T\bc + \left(\theta_2 + \btheta_3^T\ba\right) \left[\beta_0 + \bbeta_1^T\ba +\bbeta_{2}^T\bc\right] \\
\E_Y(Y_{\ba^*M_{\ba^*}} | \bC ) &=& \theta_0 + \btheta_1^T\ba^* +  \btheta_4^T\bc + \left(\theta_2 + \btheta_3^T\ba^*\right) \left[\beta_0 + \bbeta_1^T\ba^* +\bbeta_{2}^T\bc\right] \\
\end{eqnarray*}

Thus, 
\vspace*{-4mm}\begin{eqnarray*}
CDE(m) &=& \E_Y(Y | \bA=\ba, M=m, \bC = \bc) - \E_Y(Y | \bA=\ba^*, M=m, \bC = \bc) \\
	&=&\left(\btheta_1^T+ \btheta_3^T m \right)\left(\ba-\ba^*\right)  \\
NDE &=& \int_{\bc} \E_Y(Y_{\ba M_{\ba^*}} | \bC )-  \E_Y(Y_{\ba^*M_{\ba^*}} | \bC ) dP_{\bC}(\bc) \\
	&\approx& \E_Y(Y_{\ba M_{\ba^*}} | \bCbar )-  \E_Y(Y_{\ba^*M_{\ba^*}} | \bCbar )\\
	&=& \theta_0 + \btheta_1^T\ba +  \btheta_4^T\bcbar + \left(\theta_2 + \btheta_3^T\ba\right) \left[\beta_0 + \bbeta_1^T\ba^* +\bbeta_{2}^T\bcbar\right] - \\
	&&\left(\theta_0 + \btheta_1^T\ba^* +  \btheta_4^T\bcbar + \left(\theta_2 + \btheta_3^T\ba^*\right) \left[\beta_0 + \bbeta_1^T\ba^* +\bbeta_{2}^T\bcbar\right] \right)\\
	&=& \btheta_1^T\left(\ba-\ba^*\right) + \btheta_3^T \left(\ba-\ba^*\right)\left[\beta_0 + \bbeta_1^T\ba^* +\bbeta_{2}^T\bcbar\right] \\
NIE &=&  \int_{\bc} \E_Y(Y_{\ba M_{\ba}} | \bC )-  \E_Y(Y_{\ba M_{\ba^*}} | \bC ) dP_{\bC}(\bc) \\
	&\approx& \E_Y(Y_{\ba M_{\ba}} | \bCbar )-  \E_Y(Y_{\ba M_{\ba^*}} | \bCbar )\\
	&=& \theta_0 + \btheta_1^T\ba +  \btheta_4^T\bcbar + \left(\theta_2 + \btheta_3^T\ba\right) \left[\beta_0 + \bbeta_1^T\ba +\bbeta_{2}^T\bcbar\right] - \\
	&& \left(\theta_0 + \btheta_1^T\ba +  \btheta_4^T\bcbar + \left(\theta_2 + \btheta_3^T\ba\right) \left[\beta_0 + \bbeta_1^T\ba^* +\bbeta_{2}^T\bcbar\right]\right) \\
	&=& \left(\theta_2 + \btheta_3^T\ba\right)\left[ \bbeta_1^T\left(\ba-\ba^*\right) \right]
\end{eqnarray*}

\noindent When considering traditional approaches to model the outcome, we do not include exposure-mediator interactions in (\ref{outmodelLM}). We therefore model the outcome as:
\begin{eqnarray}  \label{outmodelLMtrad}
	\E[Y] &=& \gamma_0 + \bgamma_1^T\bA + \gamma_2M +  \bgamma_3^T\bC.
\end{eqnarray}

\noindent We estimate the traditional mediation effects for an exposure mixture as:
\begin{eqnarray*}
NDE &=&  \bgamma_1^T\left(\ba-\ba^*\right), \\
NIE&=& \theta_2  \bbeta_1^T \left(\ba-\ba^*\right). \\
\end{eqnarray*}

For both the linear and traditional methods, we model the total effect (TE) by:
\begin{eqnarray} \label{TEmodelLM}
	\E[Y] &=& \alpha_0 + \balpha_1^T\bA  + \balpha_2^T\bC,
\end{eqnarray}

and estimate the TE as: $\balpha^T\left(\ba-\ba^*\right)$.\\


\newpage

\begin{table}[h]
\singlespacing
\centering
\begin{tabular}{|c|l|l|l|}
  \hline\hline
Scenario & $h_M$  & $h_Y$  \\ 
  \hline\hline
    1 & $0$ & $h_\text{2,lin}(Mn,BL)$ \\ 
    2 & $0$ & $h_\text{2,quad}(Mn,BL)$ \\ 
    3 & $h_\text{1,quad}(Mn)$ & $h_\text{1,quad}(Mn)$ \\ 
    4 & $h_\text{1,quad}(Mn)$ & $h_\text{1,quad}(BL)$\\ 
    5 & $h_\text{1,quad}(Mn)$ & $h_\text{2,lin}(Mn,BL)$\\ 
    6 & $h_\text{2,quad}(Mn,Pb)$ & $h_\text{2,quad}(Mn,BL)$ \\
 \hline
\end{tabular}
 \caption[]{ Exposure-mediator and exposure-mediator-response surfaces used for data generation in our simulations.  The metals and BL indicate which elements are taken to have a true effect on the mediator and outcome. When $h_M=0$, no metals have a true effect on the mediator. The true NIE is null for scenarios 1 through 3, and the true NDE is null for scenario 4. The linear method is correctly specified for scenario 1 since $h_M(\cdot)$ and $h_Y(\cdot)$ are linear functions. Since all scenarios include an interaction term, the traditional method is misspecified for all scenarios. Graphical representation of the $h(\cdot)$ functions are depicted in Figure \ref{hfunc}.}
 \label{simscenarios}
\end{table}


\end{document}